\documentclass{aa}  

\usepackage{graphicx}
\usepackage{txfonts}
\usepackage{color}

\begin{document}

   \title{Individual power density spectra of {\em Swift} gamma--ray bursts\thanks{Tables~\ref{tbl-1} to \ref{tbl-4} are only available in electronic form at the CDS via anonymous ftp to cdsarc.u-strasbg.fr (130.79.128.5) or via http://cdsweb.u-strasbg.fr/cgi-bin/qcat?J/A+A/}}


   \author{C.~Guidorzi\thanks{guidorzi@fe.infn.it}
          \inst{1}
          \and
          S.~Dichiara\inst{1,2}
          \and
          L.~Amati\inst{3}
          }

   \institute{Department of Physics and Earth Sciences, University of Ferrara, via Saragat 1,
     I-44122, Ferrara, Italy\\
     \and
     ICRANet, P.zza della Repubblica 10, I-65122, Pescara, Italy\\
     \and
     INAF -- Istituto di Astrofisica Spaziale e Fisica Cosmica di Bologna, via Gobetti 101, I-40129, Bologna, Italy
   }


 
   \abstract
       {Timing analysis can be a powerful tool with which to shed light on the still obscure emission physics and geometry
         of the prompt emission of gamma--ray bursts (GRBs).
         Fourier power density spectra (PDS) characterise time series as stochastic processes and can be used to
         search for coherent pulsations and, more in general, to investigate the dominant variability timescales
         in astrophysical sources. Because of the limited duration and of the statistical
         properties involved, modelling the PDS of individual GRBs is challenging,
         and only average PDS of large samples have been discussed in the literature thus far.}
       {We aim at characterising the individual PDS of GRBs to describe their variability in terms of a stochastic
         process, to explore their variety, and to carry out for the first time a systematic search for periodic
         signals and for a link between PDS properties and other GRB observables.}
       {We present a Bayesian procedure that uses a Markov chain Monte Carlo technique and apply it to study the
         individual power density spectra of 215 bright long GRBs detected with the {\em Swift} Burst Alert Telescope
         in the 15--150~keV band from January 2005 to May 2015.
         The PDS are modelled with a power--law either with or without a break.}
       {Two classes of GRBs emerge: with or without a unique dominant timescale.
         A comparison with active galactic nuclei (AGNs) reveals similar distributions of PDS slopes. Unexpectedly,
         GRBs with subsecond-dominant timescales and duration longer than a few tens of seconds in the source frame
         appear to be either very rare or altogether absent.
         Three GRBs are found with possible evidence for a periodic signal at $3.0$--$3.2\,\sigma$
         (Gaussian) significance, corresponding to a multi-trial chance probability of $\sim$1\%.
         Thus, we found no compelling evidence for periodic signal in GRBs.}
       {The analogy between the PDS of GRBs and of AGNs could tentatively indicate similar stochastic processes that rule
         BH accretion across different BH mass scales and objects.
         In addition, we find evidence that short dominant timescales and duration are not completely independent
         of each other, in contrast with commonly accepted paradigms.}
       \keywords{ gamma-ray burst: general -- methods: statistical}

   \maketitle
%

\section{Introduction}
\label{sec:intro}
It is an observationally established fact that long gamma--ray bursts (GRBs), or at
least most of them, are associated with the collapse of some type of
hydrogen--stripped massive stars \citep{Woosley06,Hjorth13rev}.

The most important and open questions, however, concern the nature of the ejecta and their
magnetisation degree, where and how energy dissipation takes place, and the
radiation mechanism(s). All these questions are intertwined so that
explaining all the observed properties in a consistent picture is challenging.

Generally, several types of dissipation processes have been proposed: i) the so--called
internal shock (IS) model \citep{Rees94,Narayan92},
in which the kinetic energy of a baryon load is dissipated into gamma rays through
shocks that take place far away from the Thomson photosphere, at $10^{14}$--$10^{16}$~cm;
ii) the photospheric models, in which the dissipation occurs near the photosphere,
and where a Planckian spectrum is modified by some heating and Compton scattering
\citep{Rees05,Peer06b,Thompson07,Derishev99,Rossi06,Beloborodov10,Titarchuk12,Giannios08,Meszaros11,Giannios12};
iii) magnetised ejecta ($\sigma=B^2/4\pi\Gamma\rho c^2 > 1$) dissipate their energy at
distances similar to those of IS \citep{LyutikovBlandford03,ICMART,BeniaminiGranot15}.

The study of time variability through Fourier power density spectra (PDS; see
\citealt{Klis89,Vaughan13} for reviews) can help to constrain the energy dissipation in GRBs
as a stochastic process and also the dissipation region \citep{Titarchuk07}.
Owing to the statistical noise of counting photons in the detector \citep{Klis89}, in the GRB
literature only average PDS of long GRBs were studied
so far \citep{Beloborodov00a,Ryde03,Guidorzi12,Dichiara13a}.
The average PDS can be modelled with a power--law extending from a few
$10^{-2}$ to $\sim1$~Hz. Power--law indexes vary from $\sim1.5$ to $\sim2$ depending on the
energy passband, with steeper PDS corresponding to softer photons. 
The smooth behaviour of the PDS averaged over a large number of GRBs facilitates modelling. At the same time, it prevents studying the individual
properties of GRBs, whose light curves are non--stationary, short--lived time series.
Specifically, we cannot search for either periodic, quasi-periodic signals, or
correlations with other intrinsic properties (e.g. related to the energy spectrum).
The need to overcome these limitations calls for a proper statistical treatment of
individual PDSs.

In this work we develop a Bayesian Markov chain Monte Carlo technique
that builds
upon the procedure outlined by \citet[hereafter V10]{Vaughan10}.
We then apply it to model the PDS of individual GRBs and study the statistical
properties of an ensemble of GRBs detected in the 15--150~keV energy band with the
Burst Alert Telescope (BAT; \citealt{Barthelmy05}) onboard the {\em Swift} satellite \citep{Gehrels04}.
We chose the {\em Swift} catalogue to ensure a large homogeneous sample; in addition,
thanks to the large portion ($\sim$~30\%) of GRBs with measured redshift,
we can access a correspondingly larger set of intrinsic properties.
The same technique was recently adopted for studying a selected sample of bright
short GRBs \citep{Dichiara13b}.
A very similar approach was used for studying the outbursts from magnetars \citep{Huppenkothen13}
and for searching them for quasi--periodic oscillations (QPOs; \citealt{Huppenkothen14,Huppenkothen14b}).

The advantage of studying individual vs. averaged PDS is threefold: i)
we directly probe the variety of stochastic processes taking place
during the gamma--ray prompt emission; ii) we search for possible
connections between PDS and other key properties of the prompt
emission, such as the intrinsic (i.e. source rest-frame) peak
energy $E_{\rm p,i}$, and the isotropic--equivalent radiated energy,
$E_{\rm iso}$, involved in the eponymous correlation \citep{Amati02};
iii) we search for occasional features emerging from the PDS continuum, such
as coherent pulsations or QPOs, which, if any, would be completely washed out by
averaging the PDS of many different GRBs.
In a companion paper \citep[hereafter D16]{Dichiara16} we focus on the
$E_{\rm p,i}$--PDS correlation and its theoretical implications for a large number
of GRBs with known redshift detected by several past and present spacecraft.

The paper is organised as follows: the data selection and
analysis are described in Sect.~\ref{sec:data}. Section~\ref{sec:res}
reports the results, which are discussed in Sect.~\ref{sec:disc}.
The description of the technique adopted for the PDS modelling
is reported in Appendix~\ref{sec:pds_fit_app}.
Uncertainties on the best--fitting parameters are given at 90\%
confidence for one parameter of interest, unless stated otherwise.

\section{Data analysis}
\label{sec:data}

\subsection{Data selection}
\label{sec:data_sel}
From an initial sample of 961 GRBs detected by BAT from January 2005 to May 2015
we selected those whose time profiles were entirely covered in burst mode, that is,
with the finest time resolution available. As a consequence, GRBs discovered offline
were excluded. For the surviving 877 GRBs we extracted mask-weighted,
background-subtracted light curves with a uniform binning time of 4~ms in two
separate energy channels, 15--50 and 50--150~keV, and in the total passband
15--150~keV.
As in \citet[hereafter G12]{Guidorzi12}, we did not split the full passband
into finer energy bands to ensure a good signal--to--noise ratio (S/N) for the light
curves of most GRBs.

Light curves were extracted from the corresponding BAT event files.
The latter were processed with the HEASOFT package (v6.13) following the
BAT team threads.\footnote{
\url{http://swift.gsfc.nasa.gov/docs/swift/analysis/threads/bat_threads.html}}
Mask-weighted light curves were extracted using the ground-refined
coordinates provided by the BAT team for each burst through the
tool {\tt batbinevt}. We built the BAT detector quality map of each GRB
by processing the next earlier enable or disable map of the detectors.
Light curves are expressed as background-subtracted count rates per
fully illuminated detector for an equivalent on-axis source.

For each of these GRBs we calculated the PDS following the
procedure described in G12. The final selection was made by
imposing a threshold of S/N$\ge30$ on the total 15--150~keV fluence
collected in the time interval selected for the PDS extraction.
The choice for this particular value for the S/N threshold is
explained in Sect.~\ref{sec:res}.
From the 218 GRBs that remained at this stage, we selected the long
bursts by requiring $T_{90}>3$~s, where $T_{90}$ were taken from the second
BAT catalogue \citep{Sakamoto11} or from the corresponding BAT-refined
circulars for the most recent GRBs not included in the catalogue.
This requirement on the GRB duration excluded the only three short bursts
that had fulfilled the S/N criterion: 051221A, 060313, and 100816A.
We verified that no short GRB with extended emission \citep{Norris06,Sakamoto11}
with $T_{90}>3$~s slipped into the long-duration sample.

\subsection{PDS calculation}
\label{sec:pds_calc}
Like G12, we determined the $T_{7\sigma}$ interval, which spans from the first and
the last time bins whose rates exceed by $\ge7\,\sigma$ the background.
For most {\em Swift} GRBs its duration is very similar to that of $T_{90}$.
The PDS was then calculated on a $3\times T_{7\sigma}$ long interval with the same
central time as the $T_{7\sigma}$ sample, as the result of a trade--off between
covering the overall GRB profile and optimising the S/N (G12).
Table~\ref{tbl-1} reports the time intervals used for each of the 215 GRBs
along with the corresponding $T_{7\sigma}$ and $T_{90}$. The latter values
were taken either from the official BAT catalogue \citep{Sakamoto11} when
available, or otherwise from the BAT refined GCN circulars.
Reporting both the selected time intervals and $T_{7\sigma}$ is not redundant because for a fraction of GRBs $3\times T_{7\sigma}$ was longer than
the available data in burst mode.

The Leahy normalisation was adopted for the PDS, in which the white-noise level that is due
to uncorrelated statistical noise has a value of 2 for pure Poissonian noise \citep{Leahy83}.

\begin{table*}
\centering
\caption{Sample of 215 GRBs. The PDS is calculated in the time interval reported. }
\label{tbl-1}
\begin{tabular}{lrrrrclc}
\hline
GRB & $t_{\rm start}^{\mathrm{(a)}}$ & $t_{\rm stop}^{\mathrm{(a)}}$ & $T_{7\sigma}$ & $T_{90}$ & Cat $T_{90}^{\mathrm{(b)}}$ & $z$ & $z$ Ref$^{\mathrm{(c)}}$\\
    & (s)                             & (s)                            & (s)          & (s) &    & &\\
\hline
050117  & $ -200.318$ & $  302.658$ & $  205.312$ & $  166.65$ & S11 &  -      & - \\
050124  & $   -3.624$ & $    6.360$ & $    3.328$ & $    3.93$ & S11 &  -      & - \\
050128  & $  -35.112$ & $   51.096$ & $   28.736$ & $   28.00$ & S11 &  -      & - \\
050219A & $  -30.200$ & $   46.984$ & $   25.728$ & $   23.84$ & S11 & $0.2115$& (42)\\
050219B & $  -98.504$ & $  111.544$ & $   70.016$ & $   28.74$ & S11 &  -      & - \\
050306  & $ -188.128$ & $  302.752$ & $  186.816$ & $  158.43$ & S11 &  -      & - \\
050315  & $ -177.200$ & $  195.664$ & $  124.288$ & $   95.57$ & S11 & $1.95$  & (1) \\
050326  & $  -49.472$ & $   69.376$ & $   39.616$ & $   29.44$ & S11 &  -      & - \\
050401  & $  -42.568$ & $   64.376$ & $   35.648$ & $   33.30$ & S11 & $2.90$  & (1)\\
\hline
\end{tabular}
\begin{list}{}{}
\item[$^{\mathrm{(a)}}$]{Times are given with reference to the BAT trigger time.}
\item[$^{\mathrm{(b)}}$]{S11=\citet{Sakamoto11}; GCN= {\em Swift}-BAT-refined GCN circulars.}
\item[$^{\mathrm{(c)}}$]{(1) \citet{Hjorth12}.}
\end{list}
\end{table*}

For each light curve we initially calculated the PDS by keeping the original
minimum binning time of 4~ms, which corresponds to a Nyquist frequency of
$125$~Hz. After we ensured that no high--frequency ($f\ga 10$~Hz) periodic feature
stood out from the continuum, we decided to cut down the long computational
time demanded by Monte Carlo simulations by binning up the light curves to
$32$~ms, equivalent to a Nyquist frequency $f_{\rm Ny}=15.625$~Hz.
We did not adopt the potentially alternative approach of binning up along
frequency after we noted that the corresponding distribution of power at
high frequencies significantly deviated from the expected $\chi^2_{2M}$,
where $M$ is the re--binning factor \citep{Klis89}.
We investigated the cause for this and found out that there are two independent
reasons: one is related to the question of the correlated power at low frequencies and
is discussed in the shortcomings of our technique (Sect.~\ref{sec:limits_issues});
the other reason is that the mask-weighted light curve of a BAT GRB covers time intervals
during which the spacecraft rapidly slewed because of the GRB itself: consequently,
the mask-weighted rates are calculated from time-varying weights that depend on the
GRB direction relative to the spacecraft. This causes a pattern in the time series of
the uncorrelated Gaussian uncertainties of rates, and makes the resulting rate light curves
heteroscedastic.
We simulated light curves with the same pattern in the variance and obtained PDS exhibiting the
same behaviour. However, as long as the PDS is kept unbinned, this problem does not affect the results at a statistically significant level. 

Unless stated otherwise, the PDS hereafter discussed were calculated in this
way. Neither white-noise subtraction nor frequency re--binning was applied
to the original PDS.

\subsection{PDS modelling}
\label{sec:pds_fit}
Fitting the observed PDS with a given model requires knowing
the statistical distribution followed by the PDS at each frequency.
For instance, adopting standard least-squares optimisation techniques is
conceptually incorrect for an unbinned PDS
because power fluctuates according to a $\chi^2$, that is, more wildly than
a Gaussian variable. We therefore devised a proper treatment
that expands
upon the procedure outlined by V10 with some changes.
We refer to Appendix~\ref{sec:pds_fit_app} for a detailed description.

We considered two different models: the first is a simple power--law ({\sc pl}) plus
the white-noise constant,
\begin{equation}
S_{\rm PL}(f) = N\,f^{-\alpha} + B
\qquad .
\label{eq:mod_pl}
\end{equation}
For each PDS we first tried to fit the PDS with Eq.~(\ref{eq:mod_pl}) with the
following free parameters: the normalisation constant $N$ (we used $\log{N}$ as the
free parameter, see Appendix~\ref{sec:pds_fit_app}), the power--law index $\alpha$ ($>0$),
and the white-noise level $B$.

However, for many GRBs the PDS clearly showed evidence for a break in the power law.
We therefore considered a model of a power law with a break below which the slope is constant,
which is hereafter called bent power--law ({\sc bpl}) model,
\begin{equation}
S_{\rm BPL}(f) = N\,\Big[1 + \Big(\frac{f}{f_{\rm b}}\Big)^{\alpha}\Big]^{-1} + B
\qquad ,
\label{eq:mod_bpl}
\end{equation}
which is equivalent to the {\sc pl} model in the limit $f\gg f_{\rm b}$.
Below the break frequency, $f<f_{\rm b}$, the power density flattens.
We did not adopt the more complex broken power--law model such
as that of Eq.(1) in G12
that was used for the average PDS, which has an additional power--law index for the low--frequency range.
The reason is that the PDS of individual GRBs fluctuate more wildly around the model than the
average PDS of a sample of GRBs, simply because the fewer the degrees of freedom of a $\chi^2$
distribution, the higher the ratio between variance and expected value.
This makes the fit with a broken power--law very poorly constrained for most PDS of our sample because it adds an additional free parameter, five instead of four.
On top of this, a flat power density at low frequencies ($f\ll 1/T$, where $T$ is the GRB duration)
is also expected for a time-limited event, such as the light curve of a GRB.
In Appendix~\ref{sec:pds_fit_app} we discuss the plausibility of these models for
GRBs in more detail, along with the applicability limits and possible problems.

To establish whether a {\sc bpl} provides a statistically significant improvement
in the fit of a given PDS of a GRB with respect to a {\sc pl}, we used the likelihood ratio
test (LRT) in the Bayesian implementation described by V10 (see Eq.~\ref{eq:T_LRT} and
Appendix~\ref{sec:pds_fit_app} for details).
In a pure Bayesian approach, the Bayes factor should be used
as an alternative to the LRT \citep{KassRaftery95}:
model selection does not require the choice of any test statistic, but depends on the likelihood
marginalised (not maximised) over the joint prior of the parameters. The dependence on the parameters is thus
removed at the cost of computationally demanding multi-dimensional integrations and upon selection of appropriate
priors for the parameters.\footnote{The LRT based on the posterior predictive distribution naturally accounts for the
dependence of the posterior distribution over the whole parameter space; however, unlike for the Bayes factor, this is true only
for the simpler model {\sc pl} and not for the more complex {\sc bpl}.
This means that while its usage is reliable in excluding the simpler hypothesis of {\sc pl},
this is not necessarily evidence for {\sc bpl}. The case for {\sc bpl} as a plausible alternative to
{\sc pl} is supported by independent reasons discussed in Sect.~\ref{sec:issue1},
however.}

\begin{figure}
\centering
\includegraphics[width=8cm]{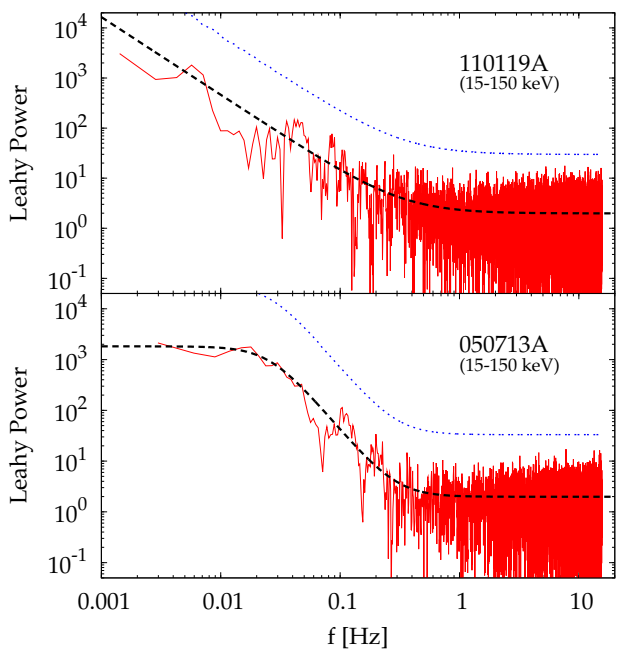}
\caption{Examples of individual PDS. Dashed lines show the corresponding
best--fitting model. Dotted lines show the $3$$\sigma$ threshold for periodic
pulsations. {\em Top}: the PDS of 110119A can be fitted with a simple {\sc pl}
model and background. {\em Bottom}: fitting the PDS of 050713A significantly
improves with a {\sc bpl} model.}
\label{fig1}
\end{figure}

For the LRT test, we accepted the {\sc bpl} model when the probability of chance improvement
was lower than 1\%.
Figure~\ref{fig1} illustrates two examples of PDS and their best--fitting models, one for
each model.
%
%

The choice of 1\% for the LRT test significance was the result of a trade--off between
type I and type II errors, also based on a set of preliminary simulated light curves.
For lower values, a number of PDS that displayed a clear--cut
break by visual inspection, and for which fitting with {\sc bpl} constrained the parameters
reasonably well, did not pass the test (too many type II errors). On the other hand, adopting
significance values higher than 1\% turned into numerous {\sc bpl}--modelled PDS with
very poorly constrained parameters (type I errors).
To calibrate the LRT threshold independently of the data, we also used two types of synthetic
light curves filled with pulses that had (i) a narrow distribution of characteristic timescales; (ii)
a broad distribution of timescales, one tenth to several ten seconds.
Peak times were assumed to be either lognormally or exponentially
distributed, in agreement with observed distributions (e.g. see \citealt{Baldeschi15} and references therein).
For the synthetic curves (i) we ensured that our procedure required {\sc bpl}, whereas no such preference
was shown for the curves (ii).
The final choice of 1\% in our sample gave only a handful of GRBs, for which we had to force the
{\sc pl} model, although the LRT test had formally rejected it. The reason was that the {\sc bpl}
parameters could not be constrained (type I errors). Just a few is also
consistent with what is expected from an equally numerous set with 1\% probability that
{\sc bpl} is mistakenly preferred to {\sc pl}.

\section{Results}
\label{sec:res}
Table~\ref{tbl-2} reports for each of the 215 selected GRBs the means and standard deviations of the
posterior distributions of the parameters of the corresponding model and the p--values of each of the
relevant statistics introduced in Appendix~\ref{sec:pds_fit_app}, in the 15--150~keV band.
Likewise, Tables~\ref{tbl-3} and \ref{tbl-4}
report the analogous information for the energy channels 15--50 and 50--150~keV, respectively.
Given that we did not re-bin the PDS along frequency and extracted the PDS out of a
single time interval, we minimised the log--likelihood of Eq.~(\ref{eq:loglik}) with $M=1$.
The number of GRBs whose PDS are best fit with {\sc bpl} is 75, 60, and 75 for the total band,
the 15--50 and the 50--150~keV channels, respectively, that is, about one-third of the sample.

To limit the effects of poorly constrained parameters in the parameters' distributions, we selected the GRBs with well-constrained power--law indexes, $\sigma(\alpha)<0.5$
(both models), and an analogous condition on $\log{(f_{\rm b})}$, $\sigma(\log{(f_{\rm b})})<0.3$,
corresponding to a factor-of-2 uncertainty on $f_{\rm b}$ for the total
band.
The sample of 215 GRBs shrank to a sub--sample of 198 with well-constrained parameters.
Within this restricted sample, the fraction of GRBs best fit with {\sc bpl} is 67/198, which is
still one-third.
%
%

\begin{table*}
\centering
\caption{Best--fitting model along with means and standard deviations of the parameters' posterior distributions
for each GRB of the sample of 215 events in the total 15--150~keV energy band.
}
\label{tbl-2}
\begin{tabular}{llrccccccr}
\hline
GRB & Model & $\log{N}$ & $\log{f_{\rm b}}$ & $\alpha$ & $B$ & $p(T_R)^{\mathrm{(a)}}$ & $p_{\rm AD}^{\mathrm{(b)}}$  & $p_{\rm KS}^{\mathrm{(c)}}$ & $N_{\rm peak}$ \\
    &       &           & (Hz)             &          &     &                            &                               &                              &               \\
\hline
050117  & {\sc PL}  & $  0.019\pm0.040$ &              NA             & $  1.347\pm0.052$ & $  1.901\pm0.029$ & $0.370$ & $0.890$ & $0.805$ & $ 15$\\
050124  & {\sc PL}  & $  0.747\pm0.146$ &              NA             & $  2.694\pm0.435$ & $  2.100\pm0.138$ & $0.653$ & $0.034$ & $0.023$ & $  2$\\
050128  & {\sc PL}  & $  0.492\pm0.049$ &              NA             & $  1.559\pm0.108$ & $  1.895\pm0.069$ & $0.147$ & $0.432$ & $0.472$ & $  7$\\
050219A & {\sc PL}  & $ -1.750\pm0.398$ &              NA             & $  2.874\pm0.361$ & $  2.014\pm0.056$ & $0.853$ & $0.730$ & $0.601$ & $  1$\\
050219B & {\sc PL}  & $  0.189\pm0.047$ &              NA             & $  1.805\pm0.070$ & $  1.856\pm0.038$ & $0.989$ & $0.350$ & $0.230$ & $  4$\\
050306  & {\sc PL}  & $ -1.249\pm0.174$ &              NA             & $  1.649\pm0.134$ & $  2.010\pm0.023$ & $0.425$ & $0.964$ & $0.956$ & $  6$\\
050315  & {\sc PL}  & $ -1.722\pm0.287$ &              NA             & $  2.116\pm0.205$ & $  2.002\pm0.026$ & $0.558$ & $0.396$ & $0.485$ & $  3$\\
050326  & {\sc BPL} & $  3.620\pm0.190$ & $ -1.060\pm0.107$           & $  2.533\pm0.123$ & $  1.875\pm0.052$ & $0.141$ & $0.699$ & $0.640$ & $  8$\\
050401  & {\sc BPL} & $  2.950\pm0.464$ & $ -1.334\pm0.281$           & $  2.312\pm0.235$ & $  2.055\pm0.053$ & $0.327$ & $0.287$ & $0.435$ & $  4$\\
\hline
\end{tabular}
\begin{list}{}{}
\item[$^{\mathrm{(a)}}$]{$p(T_R)$ is the significance associated with statistic $T_R$.}
\item[$^{\mathrm{(b)}}$]{$p_{\rm AD}$ is the significance of the Anderson--Darling test.}
\item[$^{\mathrm{(c)}}$]{$p_{\rm KS}$ is the significance of the Kolmogorov--Smirnov test.}
\end{list}
\end{table*}

We first assessed the individual and global goodness of the best--fitting models
through the distributions of the p--values associated with the Anderson--Darling (AD) and
Kolmogorov--Smirnov (KS) statistics (Appendix~\ref{sec:pds_fit_app}), finding a mean
and standard deviation of $p_{\rm AD}=0.69\pm0.26$ and $p_{\rm KS}=0.66\pm0.26$, respectively.
The distributions of both p--values are incompatible with a uniform one in the $[0:1]$
interval, being skewed towards 1. The explanation likely lies in the shortcomings of the procedure
(Sect.~\ref{sec:limits_issues}). The lowest individual p-values are a few percent, in agreement
with what is expected of a sample of 200 elements, except for 140209A,
for which both p-values were $\sim0.1$\%.
Likewise, the p--values associated with the $T_R$ statistic that
were used to search for periodic features
(see V10; Appendix~\ref{sec:pds_fit_app}) are compatible with being uniformly distributed,
as indicated by the p--value of $0.51$ of a KS test.

We investigated whether the samples that were fit with different models exhibited a significantly
different S/N: a KS test on the two S/N distributions gave an $11$\% probability of being
drawn from a common population.
We were led to adopt the threshold of S/N$\ge30$ for the sample selection because when
we had included GRBs below it in a previous attempt, the fraction of {\sc pl}--best fit GRBs
increased notably and the two S/N distributions became very different. This was interpreted
as evidence that GRBs with S/N$<30$ were just too noisy and that the stronger preference
for {\sc pl} was a mere S/N artefact.
Similarly, no evidence for a different $T_{90}$ distribution was found between the two classes.
We found a very weak indication that {\sc bpl} GRBs on average have fewer pulses per GRB,
given a KS--test probability of $7$\% that the two classes have the same number of pulses.
The number of pulses of each GRB, reported in Table~\ref{tbl-1}, had preliminarily been
determined by applying the {\sc
MEPSA} code \citep{Guidorzi15a} to the 15--150~keV band profiles.

\begin{table*}
\centering
\caption{Best--fitting model along with means and standard deviations of the parameters' posterior distributions
for each GRB of the sample of 215 events in the 15--50~keV energy band.
}
\label{tbl-3}
\begin{tabular}{llrcccccc}
\hline
GRB & Model & $\log{N}$ & $\log{f_{\rm b}}$ & $\alpha$ & $B$ & $p(T_R)^{\mathrm{(a)}}$  & $p_{\rm AD}^{\mathrm{(b)}}$  & $p_{\rm KS}^{\mathrm{(c)}}$\\
    &       &           & (Hz)             &          &     &                        &                            &                         \\
\hline
050117  & {\sc PL}  & $ -0.277\pm0.061$ &              NA             & $  1.355\pm0.062$ & $  1.911\pm0.026$ & $0.305$ & $0.749$ & $0.752$\\
050124  & {\sc PL}  & $  0.125\pm0.247$ &              NA             & $  3.012\pm0.590$ & $  1.993\pm0.133$ & $0.994$ & $0.301$ & $0.239$\\
050128  & {\sc PL}  & $ -0.059\pm0.106$ &              NA             & $  1.546\pm0.137$ & $  2.079\pm0.064$ & $0.566$ & $0.693$ & $0.541$\\
050219A & {\sc PL}  & $ -2.605\pm0.643$ &              NA             & $  3.156\pm0.506$ & $  2.032\pm0.057$ & $0.743$ & $0.563$ & $0.728$\\
050219B & {\sc PL}  & $ -0.200\pm0.077$ &              NA             & $  1.875\pm0.096$ & $  1.948\pm0.038$ & $0.895$ & $0.418$ & $0.510$\\
050306  & {\sc PL}  & $ -1.439\pm0.224$ &              NA             & $  1.529\pm0.150$ & $  1.987\pm0.023$ & $0.768$ & $0.841$ & $0.898$\\
050315  & {\sc PL}  & $ -1.708\pm0.287$ &              NA             & $  2.018\pm0.199$ & $  1.999\pm0.027$ & $0.201$ & $0.862$ & $0.774$\\
050326  & {\sc BPL} & $  3.563\pm0.353$ & $ -1.269\pm0.185$ & $  2.408\pm0.139$ & $  1.943\pm0.051$ & $0.127$ & $0.691$ & $0.606$\\
050401  & {\sc BPL} & $  2.496\pm0.341$ & $ -1.184\pm0.216$ & $  3.050\pm0.572$ & $  1.983\pm0.049$ & $0.238$ & $0.364$ & $0.242$\\
\hline
\end{tabular}
\begin{list}{}{}
\item[$^{\mathrm{(a)}}$]{$p(T_R)$ is the significance associated with statistic $T_R$.}
\item[$^{\mathrm{(b)}}$]{$p_{\rm AD}$ is the significance of the Anderson--Darling test.}
\item[$^{\mathrm{(c)}}$]{$p_{\rm KS}$ is the significance of the Kolmogorov--Smirnov test.}
\end{list}
\end{table*}

\begin{table*}
\centering
\caption{Best--fitting model along with means and standard deviations of the parameters' posterior distributions
for each GRB of the sample of 215 events in the 50--150~keV energy band.
}
\label{tbl-4}
\begin{tabular}{llrcccccc}
\hline
GRB & Model & $\log{N}$ & $\log{f_{\rm b}}$ & $\alpha$ & $B$ & $p(T_R)^{\mathrm{(a)}}$  & $p_{\rm AD}^{\mathrm{(b)}}$  & $p_{\rm KS}^{\mathrm{(c)}}$\\
    &       &           & (Hz)             &          &     &                        &                           &                         \\
\hline
050117  & {\sc PL}  & $ -0.165\pm0.051$ &              NA             & $  1.247\pm0.059$ & $  1.944\pm0.029$ & $0.424$ & $0.594$ & $0.857$\\
050124  & {\sc PL}  & $  0.861\pm0.116$ &              NA             & $  2.245\pm0.333$ & $  1.811\pm0.140$ & $0.643$ & $0.520$ & $0.460$\\
050128  & {\sc PL}  & $  0.529\pm0.046$ &              NA             & $  1.426\pm0.098$ & $  1.785\pm0.071$ & $0.054$ & $0.455$ & $0.510$\\
050219A & {\sc PL}  & $ -1.533\pm0.375$ &              NA             & $  2.684\pm0.354$ & $  1.986\pm0.055$ & $0.466$ & $0.993$ & $0.987$\\
050219B & {\sc PL}  & $  0.112\pm0.054$ &              NA             & $  1.637\pm0.072$ & $  1.897\pm0.039$ & $0.089$ & $0.972$ & $0.963$\\
050306  & {\sc PL}  & $ -1.229\pm0.203$ &              NA             & $  1.505\pm0.145$ & $  2.010\pm0.024$ & $0.237$ & $0.642$ & $0.863$\\
050315  & {\sc PL}  & $ -1.769\pm0.332$ &              NA             & $  1.849\pm0.225$ & $  1.986\pm0.026$ & $0.559$ & $0.874$ & $0.679$\\
050326  & {\sc BPL} & $  3.226\pm0.176$ & $ -0.979\pm0.110$ & $  2.437\pm0.138$ & $  1.884\pm0.053$ & $0.570$ & $0.833$ & $0.975$\\
050401  & {\sc PL}  & $  0.236\pm0.068$ &              NA             & $  1.561\pm0.106$ & $  1.986\pm0.061$ & $0.378$ & $0.971$ & $0.820$\\
\hline
\end{tabular}
\begin{list}{}{}
\item[$^{\mathrm{(a)}}$]{$p(T_R)$ is the significance associated with statistic $T_R$.}
\item[$^{\mathrm{(b)}}$]{$p_{\rm AD}$ is the significance of the Anderson--Darling test.}
\item[$^{\mathrm{(c)}}$]{$p_{\rm KS}$ is the significance of the Kolmogorov--Smirnov test.}
\end{list}
\end{table*}

\subsection{White noise}
\label{sec:res_B}
For the prior of the white-noise level $B$ we adopted a Gaussian distribution centred on 2, expected for pure
Poissonian noise, with $\sigma=0.2$, thus allowing deviations as large as $\sim10$\%.
This is a conservative approach, given the relatively little extra-Poissonian noise that affects BAT
as a result of the detector itself and of its electronics \citep{Rizzuto07}.
Another possible source of variance suppression is the detector dead time for particularly bright GRBs.
In this case the variance is suppressed by a factor $(1-\mu\,\tau)^2$, where $\mu$ is the average count
rate in an individual detector unit and
$\tau\sim100\,\mu$s\footnote{\url{http://swift.gsfc.nasa.gov/analysis/BAT_GSW_Manual_v2.pdf}} is the
BAT dead time. A 10\% dead-time suppression would imply an {\em \textup{average}} rate of several thousand counts/s
in each BAT detector module, which is way higher than the 700~events/s of the brightest events \citep{Barthelmy05}.
Therefore, our choice for the $B$ prior is very conservative, but is nonetheless useful to avoid unphysical
values.

\begin{figure}
\centering
\includegraphics[width=8cm]{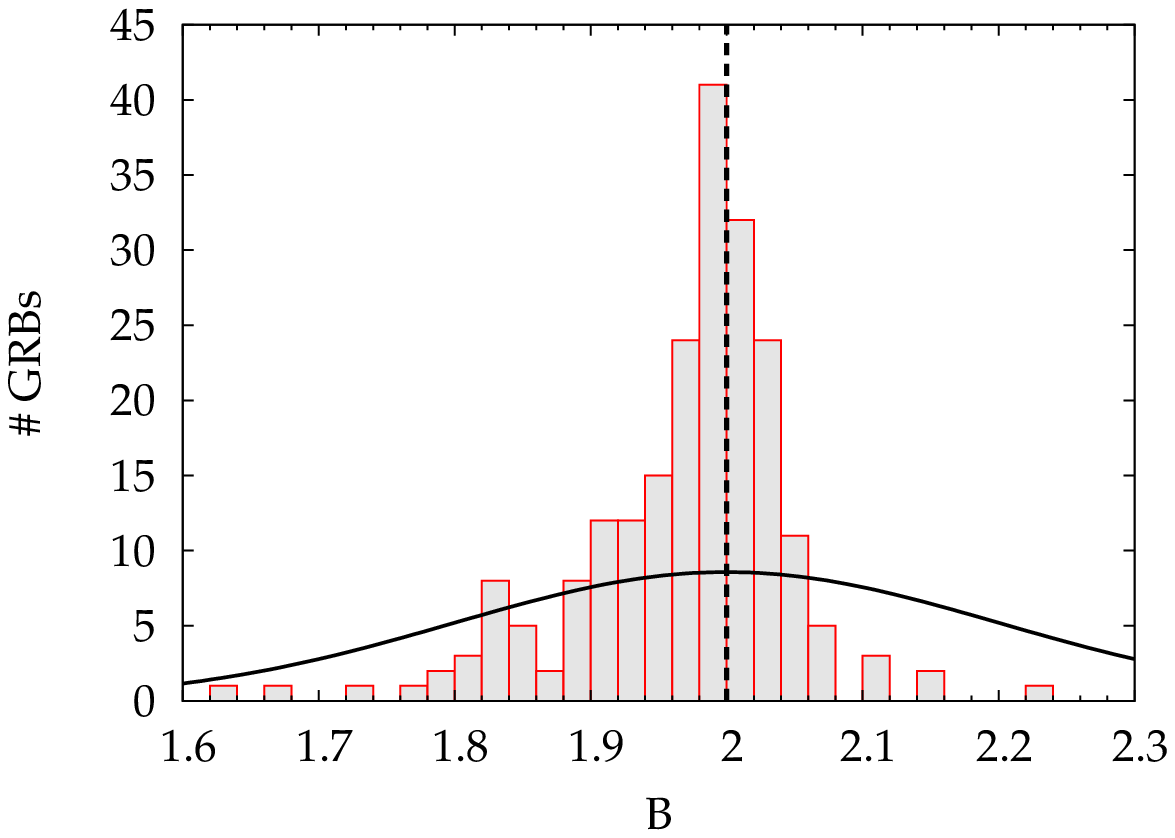}
\caption{Distribution of the white-noise level $B$.
The vertical dashed line shows the pure Poissonian case. The solid Gaussian shows the prior
distribution on $B$.}
\label{fig2_neu}
\end{figure}

The resulting distribution is shown in Fig.~\ref{fig2_neu} along with the prior distribution.
%
%
The distribution is centred on 2, as expected. Furthermore, because relatively few cases have
significantly lower values, it is skewed towards the smaller end. Most of the $B$ values are compatible
with 2 with 1 or 2 $\sigma$ ($\sigma$ is here the individual uncertainty on $B$ as obtained for each given GRB),
while for the remaining cases the dead time can hardly account for this,
as argued above. While the cause for this appears unclear, the other model parameters are insensitive
within uncertainties to it from the comparison with the results obtained in an independent run with $B=2$ fixed.
Thus, while important in itself, the choice for the $B$ prior had little effect on the other parameters of both
models. For the same reasons, we adopted the same prior for the analyses of two energy channels and 
saw no noticeable different behaviour from the total band.

\subsection{Parameter distributions}
\label{sec:res_besfitpar}
\begin{figure}
\centering
\includegraphics[width=6cm,angle=270]{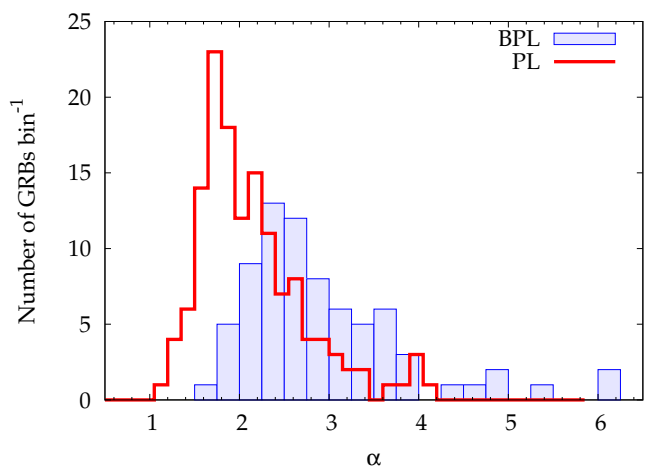}
\caption{Distribution of $\alpha$ for both models.}
\label{fig3_neu}
\end{figure}

Figure~\ref{fig3_neu} displays the power--law index distribution for both models.
As expected, on average, $\alpha$ is higher for {\sc bpl} than for {\sc pl} because
in the former model it describes the slope above the break frequency.
The median and mean values are $2.0$ and $2.1$ for the {\sc pl}, and $2.7$ and $2.9$ for
the {\sc bpl} samples.
%
%

\begin{figure}
\centering
\includegraphics[width=8.5cm]{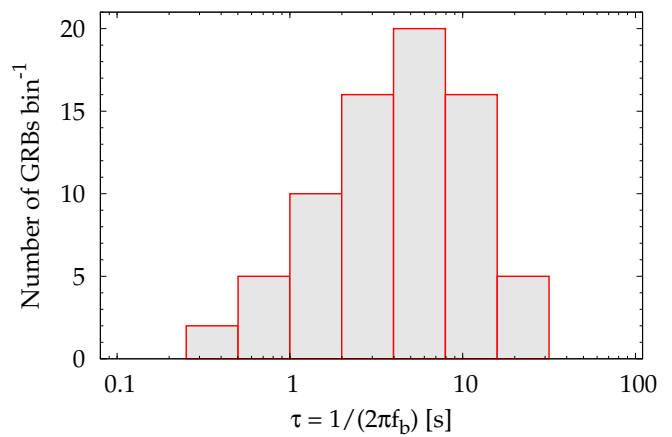}
\caption{Distribution of the dominant timescale for the {\sc bpl} model.}
\label{fig4_neu}
\end{figure}

Figure~\ref{fig4_neu} shows the distribution of the dominant timescale, defined
as $\tau=1/(2\pi f_{\rm b})$, where $f_{\rm b}$ is the break frequency as determined
by Eq.~(\ref{eq:mod_bpl}) for the {\sc bpl} model.
%
%
Although within several individual light curves we do observe sub--second
variability, when the overall variance is dominated by a
specific timescale, this mostly ranges between $0.2$ and $30$~s, with a
logarithmic average of $4.1$~s with a dispersion factor of $3$.

\subsection{Dominant timescale vs. duration}
\label{sec:res_t90_vs_tau}
\begin{figure}
\centering
\includegraphics[width=8.5cm]{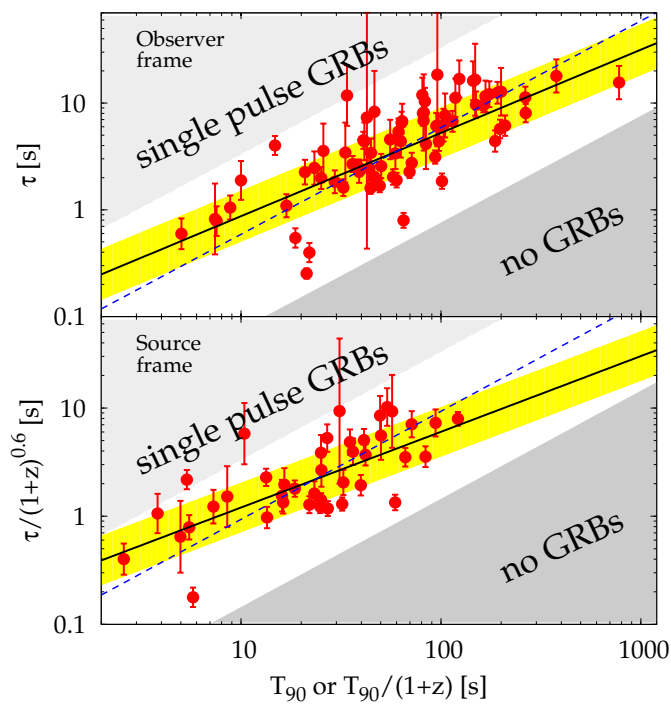}
\caption{Dominant timescale $\tau$ vs. duration $T_{90}$ for
the GRBs that are best fit with a {\sc bpl} model in both the observer's
(top panel) and in the source rest frames (bottom panel).
Solid and dashed lines show the best--fitting power law and
the best--fitting proportionality case, respectively.}
\label{fig5_neu}
\end{figure}

The 75 GRBs with a break frequency, or equivalently, with a dominant timescale
$\tau$, exhibit an interesting and unexpected property: $\tau$ is found to
correlate with the overall duration expressed in terms of $T_{90}$
(top panel of Fig.~\ref{fig5_neu}).
Its significance is reliable: $7\times10^{-15}$, $7\times10^{-15}$, and $2\times10^{-12}$
according to Pearson's linear, Spearman's, and Kendall's coefficients,
respectively.
We modelled this relation with either a simple proportionality or with a 
power law using the D'Agostini method \citep{DAgostini05} which naturally accounts
for the extrinsic scatter (i.e. additional to the measurement uncertainties
affecting each point). We conservatively assumed a 10\% uncertainty on each
value of $T_{90}$. In the former case we obtained
\begin{equation}
\tau = 10^{-1.23\pm0.06}\,T_{90}\qquad ,
\label{eq:tau_vs_t90_prop}
\end{equation}
with an extrinsic scatter $\sigma(\log{\tau})=0.25\pm0.05$ (best-fitting parameter
errors are given with 90\% confidence throughout this section). Equivalently,
on average the dominant timescale is about 20 times shorter than the overall duration,
with a dispersion of nearly $0.3$~dex.
In the latter case, modelling with a power law yields a slightly shallower dependence
of $\tau$ on $T_{90}$ than Eq.~(\ref{eq:tau_vs_t90_prop}),
\begin{equation}
\tau = 10^{-0.84\pm0.24}\ \Big(\frac{T_{90}}{{\rm 1~s}}\Big)^{0.78\pm0.13}\quad \textrm{s},
\label{eq:tau_vs_t90_pl}
\end{equation}
with an extrinsic scatter $\sigma(\log{\tau})=0.24\pm0.04$.
%
%
%

For 41 GRBs the redshift is known. For this subset we could study
the analogous relation in the GRB source rest frame. The corresponding
intrinsic quantities, denoted with subscript ${\rm i}$, were calculated
as follows: $\tau_{\rm i}=\tau/(1+z)^{0.6}$, which combines the
cosmological dilation and the narrowing of pulses with energy as
modelled by \citet{Fenimore95}; $T_{90,{\rm i}}=T_{90}/(1+z)$.
In the latter case we did not apply the narrowing of pulses to the
overall duration of the burst: this is correct especially in the
presence of waiting times. The reason is that our
sample of GRBs mostly consists of profiles with multiple pulses
interspersed with waiting times. However, we verified that the
results were not very sensitive to whether $a=0.6$ or $a=1$ is
assumed,
where the correction factor is parametrised as $(1+z)^a$.
The bottom panel of Fig.~\ref{fig5_neu} shows the result.
The correlation is still significant: the p--values are $9\times10^{-8}$,
$1\times10^{-6}$, and $8\times10^{-7}$ according to Pearson's linear,
Spearman's, and Kendall's coefficients, respectively.
Fitting the correlation with the same two models as in the observer's
frame case, we found almost identical values with larger uncertainties
that are due to the lower number of points.
\begin{equation}
\tau_{\rm i} = 10^{-1.03\pm0.08}\,T_{90,{\rm i}}
\qquad ,
\label{eq:tau_vs_t90_prop_int}
\end{equation}
with an extrinsic scatter $\sigma(\log{\tau_{\rm i}})=0.25\pm0.07$.
Modelling this with a power law, we obtain
\begin{equation}
\tau_{\rm i} = 10^{-0.62\pm0.25}\ \Big(\frac{T_{90,{\rm i}}}{{\rm 1~s}}\Big)^{0.70\pm0.18}
\quad \textrm{s},
\label{eq:tau_vs_t90_pl_int}
\end{equation}
with an extrinsic scatter $\sigma(\log{\tau_{\rm i}})=0.23\pm0.05$.

\begin{figure}
\centering
\includegraphics[width=8.5cm]{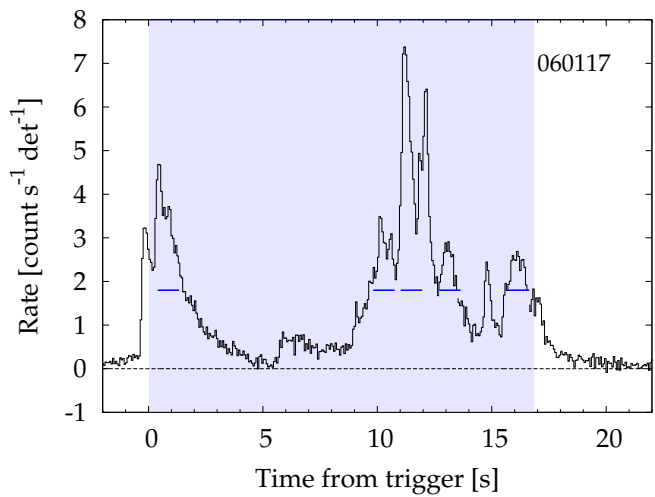}
\caption{Example of a light curve with a dominant timescale.
This is the 15--150~keV profile of 060117, whose PDS is
fit with {\sc bpl} with $\tau=1.0$~s. Horizontal solid
bars are as long as $\tau$ and emphasise the relevance
of this timescale within the overall variability.
The shaded area shows the $T_{90}=16.9$--s interval.}
\label{fig6_neu}
\end{figure}

The average ratio between $T_{90}$ and $\tau$ is higher than one order of magnitude.
This rules out that the correlation has an obvious origin.
Even for the few GRBs in our sample whose light
curves consist of a single featureless pulse, no dominant
timescale is identified. The reason is that the time
interval chosen for the PDS calculation is not long enough
to associate the break with the pulse duration itself
that appears in the PDS so as to demand a {\sc bpl} instead of
{\sc pl}.
To illustrate why the two quantities are connected
in no obvious way, Fig.~\ref{fig6_neu} displays the time
profile of 060117 as an example.
%
%

The top left region of the $T_{90}$--$\tau$ space appears
to be empty; we know it is populated by the obvious
cases, such as that of a single smooth pulse, in which the
only characteristic timescale is given by the pulse duration
itself. Instead, the dearth of long-lasting GRBs with a short 
dominant timescale, such as $T_{90}/\tau\gg 1,$ has
no explanation. To ensure that this is not an artefact of our
procedure, we constructed a synthetic light curve by replicating
and appending a real GRB profile with a short dominant timescale,
so as to obtain an arbitrarily long GRB. As a result,
our procedure did identify the same short dominant timescale
within uncertainties, whereas the duration increased by
construction. This fake GRB lay in the empty
region. This rules out any selection bias in our procedure
against this type of GRBs, and it raises the question as to
why they are rarely seen.
In conclusion, instead of a true correlation between $\tau$ and
$T_{90}$, the property that demands an explanation
is the observed dearth of short--timescale-dominated long--$T_{90}$ GRBs.

It might be wondered whether this holds for so-called ultra events
\citep{Levan14,Gendre13,Stratta13,Virgili13,Zhang14c,Boer15}.
We therefore report our results on 130925A \citep{Evans14,Piro14},
although {\em Swift}-BAT data do not cover it in its entirety.
Because of this, we did not include it
in our selected sample from which the above ensemble properties have
been drawn. In particular, the dominant timescale found by us, $\sim22$~s
(Table~\ref{tbl-1}), was inevitably extracted from the first $10^3$~s and is therefore
not descriptive of the several-ks-long profile, so the issue remains unsettled.

\begin{figure}
\centering
\includegraphics[height=8.0cm,angle=270]{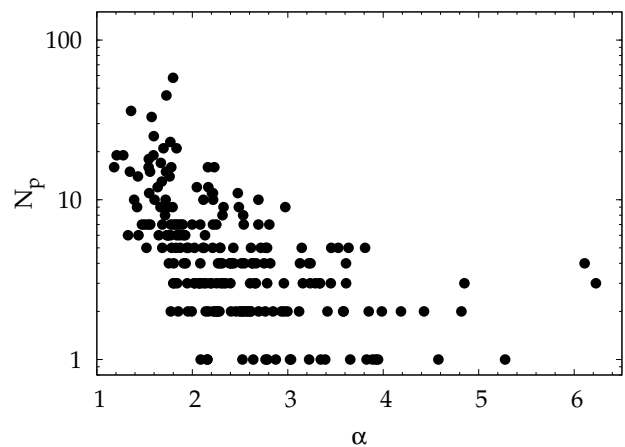}
\caption{Number of pulses as a function of the PDS slope $\alpha$.
GRBs that consist of a large number of pulses are more likely to
exhibit a shallower PDS.}
\label{fig7_neu}
\end{figure}

In Fig.~3 of D16 we illustrated the difference between the two groups
of PDS that are best fit with either {\sc bpl} or {\sc pl} and the meaning of
dominant timescale, wherever one exists. In the most general case,
the PDS concerns a light curve that is the result of superposing a number of pulses with different timescales. Whenever the total
variance is mostly dominated by some specific timescale, this shows itself
as the break in the PDS, which is best fit with {\sc bpl}.
Otherwise, when several different timescales have similar weights in the
total variance, the resulting PDS exhibits no clear break and appears to be
remarkably shallower ($\alpha_{\rm pl}\la 2$) than that of individual pulses
($\alpha_{\rm bpl}>2$). If this interpretation is correct, profiles of GRBs
with $\alpha_{\rm pl}\la 2$ on average should have more pulses.
This is indeed the case, as shown in Fig.~\ref{fig7_neu}.
%
%

\subsection{PDS and peak energy $E_{\rm p,i}$}
\label{sec:Ep_vs_alpha}
We searched for possible relations between the properties of the
PDS and intrinsic quantities of the prompt emission. In particular,
from the sample with known redshift, we selected the GRBs with
a well-constrained peak energy $E_{\rm p,i}$ of the time--averaged
energy spectrum $E\,F(E)$. 

\begin{figure}
\centering
\includegraphics[width=8.5cm]{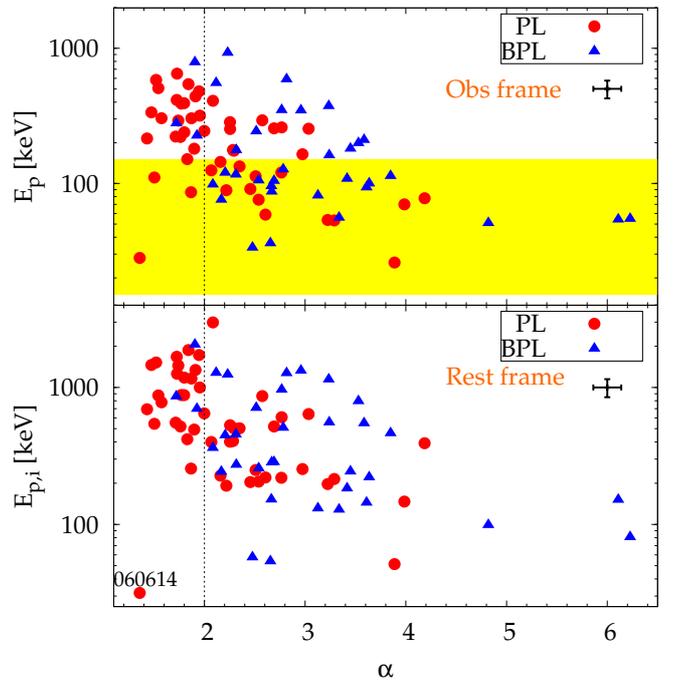}
\caption{{\em Top panel}: observed peak energy $E_{\rm p}$ vs. the power--law
index $\alpha$ of the PDS (15--150~keV) for a sample of GRBs with
known redshift and well-constrained time--averaged energy spectrum.
The shaded area highlights the BAT energy passband.
{\em Bottom panel}: same plot in the intrinsic plane, i.e. where the peak
energy $E_{\rm p,i}$ refers to the GRB comoving frame.
Circles (triangles) correspond to {\sc pl} ({\sc bpl}) model.
Median errors are shown (top right).
The dashed line shows the case $\alpha=2$.} 
\label{fig8_neu}
\end{figure}

No correlation between $E_{\rm p,i}$ and $\tau_{\rm i}$ was found for
the subsample of GRBs with both observables.
Conversely, we found a link between the PDS power--law index $\alpha$ for
both models and $E_{\rm p,i}$ as shown in Fig.~\ref{fig8_neu} (bottom panel),
which displays the two quantities for a sample of 83 GRBs.
Peak energy and redshift measures for this sample were taken from D16,
where this correlation was studied in more detail
with a larger data sample from several spacecraft.
The power--law index of the PDS refers to the 15--150~keV light curve.
On average, GRBs with higher peak energies exhibit lower PDS indexes.
The p--values associated with Pearson's, Spearman's, and Kendall's
coefficients are $3.7\times10^{-6}$, $3.9\times10^{-7}$, and
$8.6\times10^{-7}$, respectively. These values do not account for the measurement
uncertainties. Although these centroids are already the result of
a scattering that is due to the individual uncertainties, we conservatively
estimated their effect through MC simulations: we independently scattered
each point assuming a log--normal distribution along $E_{\rm p,i}$ and
using the marginal posterior distribution obtained for $\alpha$ for each GRB
(Appendix~\ref{sec:pds_fit_app}). We generated 1000 synthetic sets
of 83 GRBs each and calculated the corresponding correlation coefficients.
The 90\% quantiles of the p--value distributions of the above-mentioned
correlation coefficients are $3.5\times10^{-4}$, $1.6\times10^{-5}$, and $1.5\times10^{-5}$,
respectively. Hence, the significance of the correlation according to non--parametric
tests lies in the range $10^{-5}$--$10^{-4}$.

The top panel of Fig.~\ref{fig8_neu} shows the same correlation in
the observer frame for the same sample, where the observed peak energy
$E_{\rm p}$ replaces $E_{\rm p,i}$. This is useful to determine where
$E_{\rm p}$ lies with respect to the BAT energy passband,
in which light curves were extracted to evaluate its effect.
Even considering only the GRBs whose $E_{\rm p}$ lies above the BAT
passband, there are still a few cases for both models with $\alpha>2$.
This rules out that the $E_{\rm p,i}$--$\alpha$ correlation is the result of a bias connected with the energy band in
which temporal profiles are extracted.
In the observer plane, the correlation is slightly less significant:
the same 90\% quantiles are $4.2\times10^{-4}$, $7.6\times10^{-5}$, and $7.7\times10^{-5}$ , respectively.
However, if the narrower range $\alpha<4.5$ is considered, the correlation is almost one order
of magnitude more significant in the source rest frame than in the observer frame.
The same property of a more significant correlation in the source rest frame than in the observer
frame holds for the larger sample of D16.

We refer to D16 for an exhaustive analysis of the $E_{\rm p,i}-\alpha$ correlation
and its implications in the framework of different models proposed in the literature.
As shown in D16, this correlation is found to hold and extend to GRBs detected with other
current and past experiments with far better significance.

%

\begin{figure}
\centering
\includegraphics[width=8.5cm]{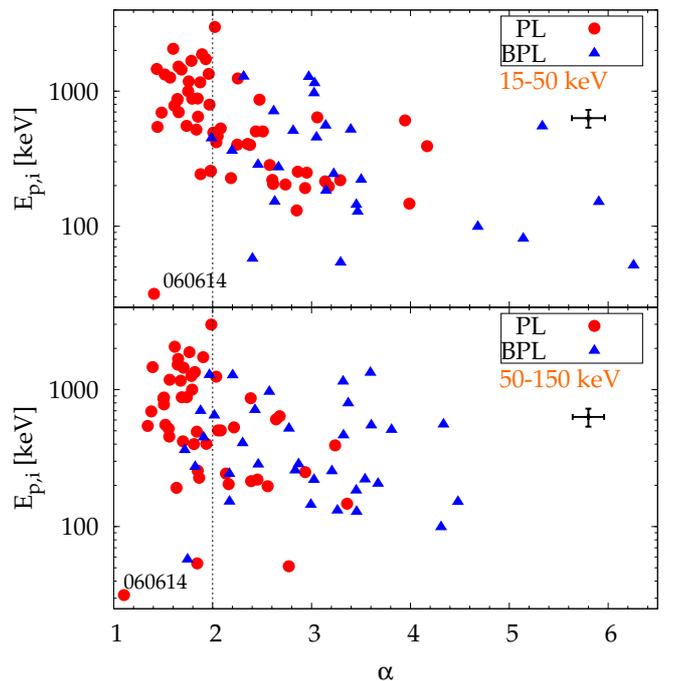}
\caption{Same as Fig.~\ref{fig8_neu}, except that $\alpha$ was obtained
for each energy channel: 15--50 (top), and 50--150~keV (bottom).}
\label{fig9_neu}
\end{figure}

We studied the same relation for the PDS of each energy channel and
show the result in Fig.~\ref{fig9_neu}.
%
%
The same correlation seen in the total energy band is also evident
in the individual energy channels with some notable differences, however.
For the lower energy channel 15--50~keV, the correlation is
even more significant than the full passband ($3\times10^{-7}$, $3\times10^{-9}$,
and $6\times10^{-9}$ significance).
Conversely, in the harder energy band the correlation is much weaker and more
scattered ($9\times10^{-4}$, $2\times10^{-4}$, and $1\times10^{-4}$ significance).

More generally, if  the sample is split into two subsets, depending
on whether it is $\alpha<2$, the correlation is no more
significant in each subset. This suggests that the correlation
might mainly be due to the existence of two different classes
of GRBs that are characterised by a shallow or a steep PDS.

To gain further insight, we performed some KS tests
on the $E_{\rm p,i}$ distributions of the two classes: in the 
total energy band case we split the sample assuming $\alpha=2$
as the dividing line. The resulting p--value of a common
$E_{\rm p,i}$ is $1.7\times10^{-5}$ ($5.9\times10^{-6}$ when 060614
is excluded). Analogous tests on the samples of the two energy
channels yield p--values of $4\times10^{-6}$ and $6\times10^{-3}$
for the 15--50 and the 50--150~keV band, respectively.
Dropping 060614, these p--values decrease to $2\times10^{-6}$
and $4\times10^{-3}$.
060614 can be treated as an anomalous source {\em \textup{for independent reasons}}:
it shares properties with both long and short GRBs: its duration and its
consistency with the $E_{\rm p}$--$E_{\rm iso}$ favour the long
classification \citep{Amati07}; the negligible temporal lag of its initial spike,
along with the significant absence of any typical SN associated to other long GRBs
\citep{Gehrels06,DellaValle06,Fynbo06}, and the possible evidence for an associated
macronova \citep{Yang15,Kisaka15,Jin15}, would place it in the short group, so that
this GRB is still singular. See D16 for a thorough discussion of 060614.

That the dividing line is around $\alpha=2$ may hide a profound
meaning in the theory of PDS formation \citep{Titarchuk07}.

\subsection{Search for periodic signal}
\label{sec:res_periodicity}
We searched the total and individual energy channel PDS
of each GRB for periodic features above a $3\sigma$ (Gaussian) significance
threshold and found none.

\begin{figure}
\centering
\includegraphics[width=8.5cm]{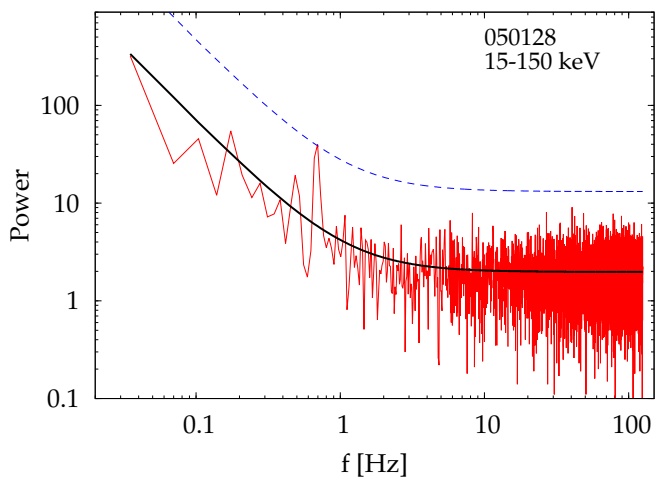}
\caption{PDS of 050128 in the 15--150~keV band. The solid (dashed)
line shows the {\sc pl} best--fitting model ($3\sigma$ level above the
continuum).}
\label{fig10_neu}
\end{figure}

\begin{figure}
\centering
\includegraphics[width=8.5cm]{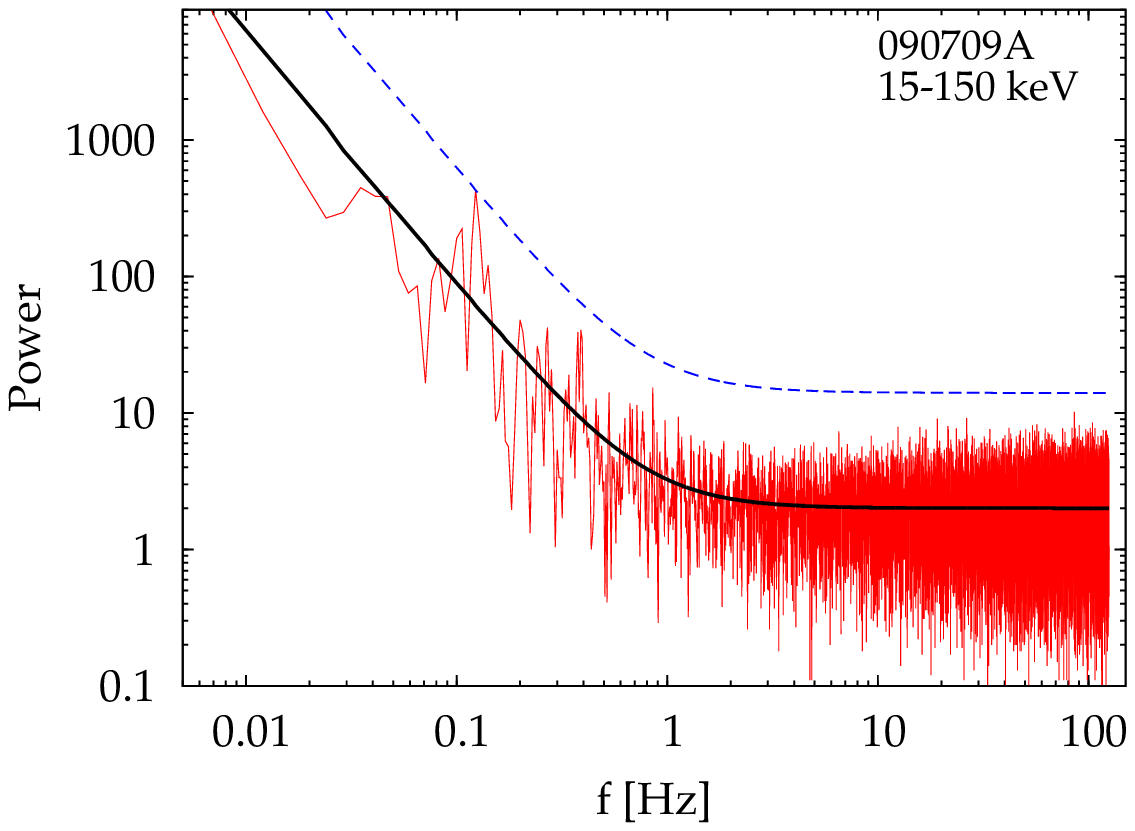}
\caption{PDS of 090709A in the 15--150~keV band. The solid (dashed)
line shows the {\sc pl} best--fitting model ($3\sigma$ level above the
continuum).}
\label{fig11_neu}
\end{figure}

\begin{figure}
\centering
\includegraphics[width=8.5cm]{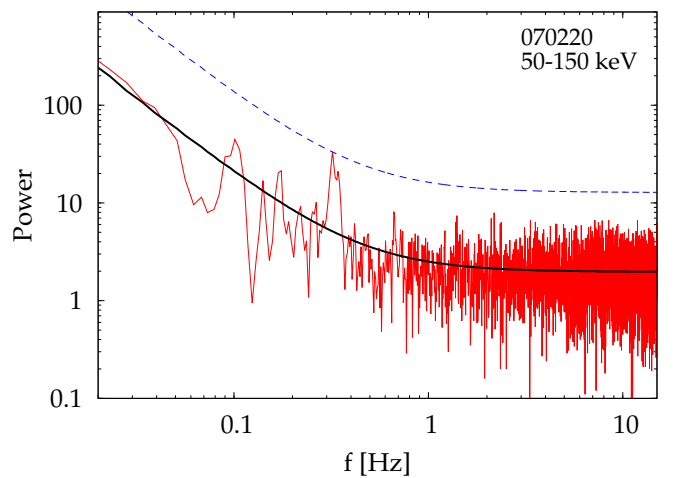}
\caption{PDS of 070220 in the 50--150~keV band. The solid (dashed)
line shows the {\sc pl} best--fitting model ($3\sigma$ level above the
continuum).}
\label{fig12_neu}
\end{figure}

We then adopted an alternative approach: we split the $3\times$$T_{7\sigma}$
interval into three equal sub--intervals, calculated the PDS for each,
and averaged out the resulting PDS. We then minimised the log--likelihood
of Eq.~(\ref{eq:loglik}) with $M=3$. The most obvious benefit of summing
different PDS is a reduced statistical noise.
Performing the search for periodic pulsations over these average PDS for
each individual GRB in the total passband and in the two energy
channels allowed us to pick up three GRBs with $>3\sigma$ significance.
In the total passband 050128 (Fig.~\ref{fig10_neu}) and 090709A
(Fig.~\ref{fig11_neu}) show a feature at $f=0.696\pm0.035$~Hz
and at $f=0.123\pm0.006$~Hz, respectively. 070220 shows an excess in the 50--150~keV
band at $f=0.321\pm0.006$~Hz (Fig.~\ref{fig12_neu}).%
%
%
%
The significance of each feature was determined by the p--value associated
with the $T_R$ statistic in each case: the three corresponding p--values 
amount to $1.4\times10^{-3}$, $2.3\times10^{-3}$, and $2.6\times10^{-3}$
for 050128, 090709A, and 070220, respectively. In units of Gaussian
$\sigma$'s, they correspond to $3.2$, $3.0$, and $3.0$.
For 090709A similar searches gave analogous results
\citep{DeLuca10,Cenko10}.
Because of the reasons explained in Appendix~\ref{sec:pds_fit_app}, the
significance accounts for the multi-trial search over the whole
range of explored frequencies in each individual spectrum.
The question we addressed instead concerns the probability that at least
three GRBs out of 200 show features equal to or more significant than
the observed ones by chance. To this aim, we took the smallest significance,
that is, $p=1.4\times10^{-3}$, as the success probability of a single trial.
We used a binomial distribution with $N=200$ trials and determined
the probability of having $n\ge3$ successful events by chance,
which is $0.2$\%. We might question whether
this underestimates the true value. Instead, when we took the highest
p--value as the success probability for the single trial, that
is, $p=2.6\times10^{-3}$, the resulting probability for the multi-trial
was a mere 1\%. Thus, the true value lies between $0.2$ and 1\%.
This value does not allow us to make a strong statement about the
evidence for  coherent pulsations in some GRBs.
However, there is a further, more subtle but nonetheless crucial
caveat we did not mention so far, and which lies in the procedure
with which we calculated the PDS for this search. Unlike for the previous parts of
the present investigation, here we sliced the light curves into
sub-intervals and averaged the PDS of each of them out.
This is common practice for steady sources because it implicitly
relies on the assumption of an ergodic process (e.g. \citealt{Guidorzi11a}).
While this is reasonable for a steady source, this is unlikely to be the case
for the GRB signal. Therefore, in spite of the possible detection
of coherent pulsation with $\la$1\% confidence, its interpretation
is undermined by the implicit assumption of the GRB light curve
thought of as an ergodic process, which can hardly make physical sense.
We conclude that there is no unambiguous evidence for periodic features in this {\em Swift}-BAT GRB data set.

\section{Summary and discussion}
\label{sec:disc}
For the first time, a systematic analysis of individual PDS of long GRBs has been carried out
using a statistical treatment that allowed us to properly model the continuum and
search for possible periodic or QPOs in a self-consistent way.
An analogous investigation based on the same technique was carried out by us \citep{Dichiara13b}
on a sample of short GRBs, where we compared the derived distribution of the PDS slopes $\alpha$ 
with the preliminary one for a sample of long GRBs, finding no striking difference.

This technique, described in detail in Appendix~\ref{sec:pds_fit_app}, has opened up the
possibility to search for correlations between PDS and other key properties, the most
significant and remarkable of which is $E_{\rm p,i}$--$\alpha$.
In D16 we thoroughly studied this correlation over an enlarged sample of GRBs detected with
different spacecraft and discussed its physical implications in the context of some
prompt emission models. In particular, we linked the PDS slope to the relative strength
of the fast component (timescale $\lesssim 1$~s) in the light curve: when this is clearly
present, the PDS is shallow ($\alpha\lesssim2$) and shows no break. By contrast,
when it is either weak or missing, the PDS is steeper ($\alpha>2$) with or without a break in the
range $0.01$--$1$~Hz.
While a KS test between the $T_{90}$ distributions of the two groups of PDS models reveals
no difference (p-value of 65\%), the {\sc bpl} group on average possibly has fewer peaks/GRB
(KS p-value of $7$\%), as also shown in Fig.~\ref{fig7_neu} considering that they have higher
$\alpha$s than the {\sc pl} group.
Almost all of GRBs with $N>10$ peaks have PDS slopes $\alpha<2$.
This agrees with our interpretation of the PDS: GRBs with many peaks are more likely to cover
a broader range of timescales, so that the resulting PDS is shallow and has no dominant
timescale (see Fig.~3 in D16).
The interpretation in terms of two independent components
in GRB light curves is in accord with the results of previous investigations based on
different techniques \citep{Shen03,Vetere06,Gao12}.

As we discussed in D16, in the context of some of the GRB prompt emission models the observed variability
tracks the intrinsic inner engine behaviour. If the engine is a newly formed black hole (BH) accreting
from a disk, as predicted in the collapsar model \citep{Woosley93,MacFadyen99}, it is interesting
to compare our results with the PDS properties of other astrophysical sources that are known to be powered
by accreting BHs. Specifically, accretion around BHs is known to be characterised by some scalings
between BH mass, timescales, and accretion luminosity (relative to the Eddington limit). These scalings
hold for Galactic stellar BHs (GBHs) and for super-massive BH (SMBH) that power active galactic nuclei
(AGNs), according to the fundamental plane of BH activity \citep{Merloni03,Falcke04}.
We took the sample of AGN PDS obtained by \citet{GonzalezMartin12}, who fit them either with a {\sc pl}
or with a so-called bending power-law, which differs from our {\sc bpl} model only in the low-frequency regime,
where power scales as $f^{-1}$ instead of $f^0$ for $f\ll f_{\rm b}$. Our choice of $f^0$ for GRBs is a
consequence of the time finiteness of the GRB signal, which entails a constant power at $f\ll 1/T$,
where $T$ is the GRB duration.
On the other hand, AGNs and GBHs are stationary sources that do not undergo irreversible processes like the
sequence of events that make up the GRB phenomenon.
As such, their red noise power extends to $f\ll 1/T$, where $T$ is the duration of the observation window.
This explains why $f^{-1}$ is more appropriate for these sources than $f^0$ at $f\ll 1/T$.
We also considered the analogous distribution for a sample of magnetar bursts from \citet{Huppenkothen13},
who fitted the PDS with a similar and more general broken power-law model as an alternative to a simple {\sc pl}.
In terms of stochastic processes, the magnetar burst sample is more similar to that of GRBs because of the non-stationary
character and even shorter duration.
%
\begin{figure}
\centering
\includegraphics[height=8.5cm,angle=270]{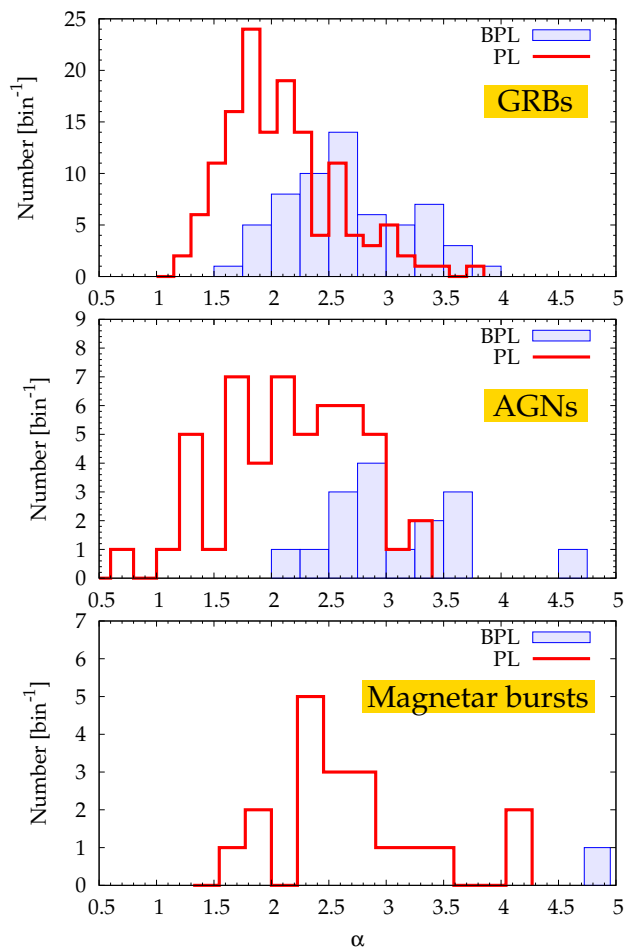}
\caption{Distribution of the PDS slope for both models (solid is {\sc pl}, filled is {\sc bpl})
for three classes of astrophysical sources: GRBs ({\em top}, this work), AGNs ({\em middle},
from \citealt{GonzalezMartin12}), and magnetar bursts ({\em bottom}, from \citealt{Huppenkothen13}).}
\label{fig13_neu}
\end{figure}
%
Figure~\ref{fig13_neu} shows the comparison between the PDS slope distribution of our GRB set,
that of AGNs, and that of magnetar outbursts.
Although all of the distributions span very similar ranges, the distributions that appear more alike are
those of GRBs and AGNs: the distributions for both models look remarkably similar, at least as far as
their respective ranges are concerned.

Furthermore, analogous trends with the photon energy are observed: PDS are shallower for harder energy channels,
as observed for AGNs \citep{GonzalezMartin12}, for GBHs \citep{Nowak99}, and for the average PDS of
GRBs \citep{Beloborodov00a,Guidorzi12,Dichiara13a}.
PDS of GBHs are generally more complicated and strongly depend on the source state: in the soft state,
they are typically modelled with a bending {\sc pl} with a high-frequency index $\alpha>2$,
whereas $\alpha$ can vary in the range between 1 and 2 \citep{Cui97,Nowak99,Remillard06rev}.
By contrast, unlike for GRBs, some AGN and GBH PDS are also characterised by QPOs. Evidence for QPOs has
recently been found in the PDS of a few magnetar bursts \citep{Huppenkothen14,Huppenkothen14b}.

Concerning the dominant timescales $\tau$ identified in the PDS, we found no correlation between $\tau$
and other intrinsic properties, such as $E_{\rm p,i}$, or $E_{\rm iso}$. In the AGN case, the timescale
corresponding to the break frequency is known to scale with the BH mass (e.g., \citealt{Markowitz03}).
In GRBs the origin of dominant timescales is likely different: on average, most low-$E_{\rm p,i}$
GRBs, which also have correspondingly low peak luminosities, have a several-second-long dominant timescale
and either weak or absent subsecond variability, as indicated by the $E_{\rm p,i}$--$\alpha$ correlation
(Figs.~\ref{fig8_neu} and \ref{fig9_neu}; see also D16). Assuming the BH scaling, we would infer higher masses
for these GRBs than for the bulk of GRBs. However, they look like they can hardly be associated
with the most massive BHs among the possible GRB progenitor candidates, given their lower luminosities.

Overall, we have to be cautious in building upon these analogies between GRBs and AGNs purely based on the
PDS slopes, firstly because of the strong non-stationary character of GRBs, secondly because the BH nature
of GRB inner engines is not yet established beyond doubt. Nonetheless, regardless
of the origin of GRB variability (see D16), the result illustrated in Fig.~\ref{fig13_neu} might
stem from a common process that rules accreting BH across different mass scales.

Concerning the presence of dominant timescales, an unexpected result from our analysis
is the absence or dearth of subsecond-dominant timescales ($\tau\lesssim1$~s) in long GRBs
($T_{90}\gtrsim10$--$20$~s in the rest frame; see bottom panel of Fig.~\ref{fig5_neu}).
To our knowledge, this type of constraint for long GRBs is not predicted in any prompt emission
model that appeared thus far in the literature. We showed that this dearth is a genuine property of
real light curves and not an artefact of our technique.
In other words, when short-timescale variability dominates the light-curve variance, the overall
duration of the prompt emission cannot last longer than a few tens of seconds.
Equivalently, very long GRBs cannot have dominant timescales as short as $\tau\lesssim1$~s, that is,
a significant fraction of the temporal power must lie in the slow ($>1$~s) component.
This unexpected constraint between short-timescale variability and overall duration of GRBs should be
taken into account in prompt emission models.

Although we found three GRBs with some evidence ($\sim 3\sigma$ significance) for coherent pulsations,
the overall multi-trial probability that these are mere statistical flukes is about 1\%, that is, not negligible.
The overall lack of unambiguous evidence for periodic signal in GRB prompt light curves does not clash with models 
in which the GRB progenitor injects energy following a periodic or quasi-periodic pattern, such as in the newly born
millisecond magnetar model \citep{Usov92,Thompson04,Metzger11}, or due to the viscous spindown of a BH
\citep{VanputtenGupta09}. The reason is that a localised release of a comparable amount of energy entails
the formation of an $e^{\pm}$ and gamma-ray fireball \citep{CavalloRees78}, which may quench the imprinted temporal pattern,
unlike what occurs for the possibly associated gravitational-wave signal \citep{Vanputten09}.
Our conclusion assumes that all GRBs are different realisations of a common stochastic process.
If we drop this assumption, we cannot exclude that the periodic patterns observed in these three GRBs are
real and that for some unknown reasons they are intrinsically different from the bulk of GRBs.


\begin{acknowledgements}
  We are grateful to the anonymous referee for insighful comments that improved the paper, and
  to Filippo Frontera, Lev Titarchuk, and Chris Koen for useful discussions.
  We acknowledge support by PRIN MIUR project on ``Gamma Ray Bursts: from
  progenitors to physics of the prompt emission process'', P.~I. F. Frontera (Prot. 2009 ERC3HT).
\end{acknowledgements}

\appendix

\section{PDS modelling}
\label{sec:pds_fit_app}
We start from the general assumption that a time series is the outcome of a stochastic process.
A broad class of processes is the result of a linear system that operates linearly on an input
process $x(t)$ and yields another stochastic process $y(t)=L[x(t)]$. One example is given by the shot
noise, which results from an input $z(t)$ of Poisson impulses,
\begin{equation}
z(t)\ =\ \sum_i \delta(t - t_i)\;,
\label{eq:shot1}
\end{equation}
where $t_i$ are Poisson points along the time axis with a given average rate $\lambda$, and the output
shot noise process is
\begin{equation}
y(t)\ =\ \sum_i h(t - t_i)\;,
\label{eq:shot2}
\end{equation}
where $h(t)$ is a given deterministic function of time, so that the linear operator is in this example 
the convolution with the chosen deterministic function, $y = h * z$.
The expected PDS $S_{yy}(\omega)$ of $y(t)$ at $\omega>0$ is given by
\begin{equation}
S_{yy}(\omega)\ =\ |H(\omega)|^2\, S_{zz}(\omega)\;,
\label{eq:shot3}
\end{equation}
where $S_{zz}(\omega)$ is the PDS of the Poisson process, so it is ruled by the characteristic $\chi^2_2$
white-noise distribution, and $|H(\omega)|^2$ is the square modulus of the Fourier transform of the deterministic function
acting like a scaling factor at any given frequency. From Eq.~(\ref{eq:shot3}) the PDS of the output process
inherits the same $\chi^2_2$ distribution, where the expected power at $\omega$ is $|H(\omega)|^2$ \citep{Israel96}.
We can complicate this further, for example, by considering shot noise derived from a generalised Poisson
process\footnote{See \citet{Papoulis}.}, in which the input process is given by an impulse
train with variable intensity, which can be itself another random variable $c_i$:
\begin{equation}
z'(t)\ =\ \sum_i c_i\,\delta(t - t_i)\;,
\label{eq:shot4}
\end{equation}
and/or consider a Poisson process with variable rate $\lambda(t)$ (e.g. for the study of shot noise in the PDS of solar
X-ray flares, \citealt{Frontera79}) and/or the combination of different deterministic functions.
Therefore, for this type of processes the coloured part of the PDS is mostly determined by the shape of the average
deterministic profile that is convolved with the point process, with possible contribution from $c_i$ when this is
characterised by correlated noise as well.

These processes yield a satisfactory description of the observed PDS of many astrophysical time series that can be treated
as stationary (or, at least, locally stationary) processes, where the deterministic function $h(t)$ describes the shape of
a single shot. Our procedure assumes that GRB time profiles can be described by a process obtained by convolving
the typical shape $h(t)$ of an individual pulse with a generalised Poisson process like that of Eq.~(\ref{eq:shot4}).
At first glance, this assumption appears to be plausible for both the deterministic and the stochastic sides of the
process: a typical pulse shape has indeed been identified in the form of a fast-rise exponential decay (FRED; \citealt{Norris96});
the sequence of pulses, treated as a point process, is compatible with a Poisson process as long as long quiescent times
are neglected \citep{Baldeschi15}. In Sect.~\ref{sec:limits_issues} we discuss the shortcomings of this assumption
and implications on the results.

For sums of independent PDS, the power in each frequency bin distributes like a $\chi^2_{2M}$,
where the degrees of freedom, $2\,M$, is given by two times $M$, that is, the number of original spectra that are summed
\citep{Klis89}.
Let $P_j$ be the observed power at frequency bin $j$ and $S_j$ its
model value. The corresponding probability density function for $P_j$ given
the expected value $S_j$ is given by
\begin{eqnarray}
p(P_j|S_j) & = & \frac{2M}{S_j}\ \chi^2_{2M}\Big(2M\,\frac{P_j}{S_j}\Big)\nonumber\\
& = & \frac{M}{S_j\Gamma(M)}\,\Big(M\,\frac{P_j}{S_j}\Big)^{M-1}\,\exp{(-MP_j/S_j)},
\label{eq:mod_pdf}
\end{eqnarray}
where $\Gamma()$ is the gamma function.

The joint likelihood function, $p(\mathbf{P}|\mathbf{S},H)$, for a given
PDS $\mathbf{P}=\{P_1, P_2, \ldots, P_{N/2-1}\}$, given a generic model $H$ with
expected values $\mathbf{S}=\{S_1, S_2, \ldots, S_{N/2-1}\}$,
is given by
\begin{equation}
p(\mathbf{P}|\mathbf{S},H) = \prod_{j=1}^{j=N/2-1} p(P_j|S_j)
\qquad ,
\label{eq:jointlik}
\end{equation}
where $N$ is the number of bins in the light curves. We excluded
the Nyquist frequency bin ($j=N/2$), since this follows a different
distribution, $\chi^2_{M}(MP_{N/2}/S_{N/2})$ \citep{Klis89}.

Maximising Eq.(\ref{eq:jointlik}) is equivalent to minimising
the corresponding un--normalised negative log--likelihood,
$L(\mathbf{P},\mathbf{S},H)$,
\begin{equation}
L(\mathbf{P},\mathbf{S},H) = \sum_{j=1}^{j=N/2-1}\ \Big(M\,\log{S_j} + M\,\frac{P_j}{S_j}
- (M-1)\,\log{P_j} \Big).
\label{eq:loglik}
\end{equation}
So far, the dependence of the joint log--likelihood in Eq.~(\ref{eq:loglik})
on model $H$ is implicit through the model values, $S_j$ (see also \citealt{BarretVaughan12}).

We determine the best--fitting model and the relative best--fitting parameters
in the Bayesian context. From the Bayes theorem, the posterior probability
density function of the parameters of a given model $H$ and for a given observed
PDS $\mathbf{P}$ is
\begin{equation}
p(\mathbf{S}|\mathbf{P},H) = \frac{p(\mathbf{P}|\mathbf{S},H)\,p(\mathbf{S},H)}{p(\mathbf{P}|H)} ,
\label{eq:bayes}
\end{equation}
where the first term in the numerator of the right-hand side of Eq.~(\ref{eq:bayes}) is
the likelihood function of Eq.~(\ref{eq:jointlik}), $p(\mathbf{S},H)$ is the prior
distribution of the model parameters, in addition to the normalising term at the denominator.

We assumed uninformative prior distributions, except for the white-noise level, for which we used
a conservative Gaussian centred on the pure Poissonian value of 2 with $\sigma=0.2$ (Sect.~\ref{sec:res_B}).
For the normalisation term $N$ we adopted Jeffrey's prior (see V10) given that it spans several decades,
whereas a flat prior was used for the remaining parameters. The question of uninformative priors is
the matter of on-going research in statistics, and the choice could depend on the specific problem.
Finding the mode of the posterior probability of Eq.~(\ref{eq:bayes}) is therefore equivalent to
minimising the negative log--likelihood (\ref{eq:loglik}).

For each PDS we adopted the following fitting procedure. First, we tried
to fit the PDS with a simple {\sc pl} model described by Eq.~(\ref{eq:mod_pl})
where the free parameters are the normalisation constant $N$, the power--law index
$\alpha$ ($>0$), and the white-noise level $B$. The logarithm of the normalisation was used
instead of $N$ itself because its posterior is more symmetric and easier to handle.

For a sizable part of our sample the PDS required the more complex model described by {\sc bpl}
(Eq.~\ref{eq:mod_bpl}). The reasons for the choice of this particular model are explained
in Sect.~\ref{sec:pds_fit}.

We adopted the Bayesian procedure presented by V10 for estimating the posterior density of
the model parameters through a Markov chain Monte Carlo (MCMC) algorithm such as
the random--walk Metropolis--Hastings in the implementation of the $R$ package
{\tt MHadaptive}\footnote{
http://cran.r-project.org/web/packages/MHadaptive/index.html.} (v.1.1-2).
V10 treated the case $M=1$, whereas we considered a more general $M\ge1$.
We started by approximating the posterior using a multivariate
normal distribution centred on the mode and whose covariance matrix is that obtained by
minimisation of Eq.~(\ref{eq:loglik}). For a given PDS, we generated $5.1\times10^4$ sets of simulated
parameters and retained one every five MCMC iterations after excluding the first 1000.
The remaining $10^4$ sets of parameters were therefore used to approximate the posterior
density. To check the quality of the fit results and search for interesting features, such
as QPO or periodic signatures superposed to the continuum spectrum, we used each set of simulated
parameters of the PDS model to generate as many synthetic PDS from the the posterior predictive
distribution. Hence for a given observed PDS, this procedure allowed us to directly calculate
$10^4$ simulated PDS and use them to infer the probability density function of all the statistics
we are interested in.

Let $\hat{S}_j$ be the model value at frequency bin $j$ obtained with the best-fit parameters at
the mode of the posterior. Following V10, we define the following quantity, $R_j = 2MP_j/\hat{S}_j$.
If the true model $S_j$ were known, $R_j$ would be exactly $\chi^2_{2M}$-distributed. However,
estimating it through $\hat{S}_j$ affects its distribution. The advantage of using the posterior
predictive distribution is that no assumption on the nature of the distribution of $R_j$ is
required when we need to determine the corresponding p--values, since its probability density
function (hereafter pdf) is sampled through
the simulated spectra and the uncertainties in the model are automatically included.
Let $\tilde{P}_{j,k}$ be the $j$-th bin power of the $k$-th simulated PDS. Correspondingly,
we also define $\tilde{R}_{j,k} = 2M\tilde{P}_{j,k}/\hat{S}_j$.
We chose three different statistics:
\begin{itemize}
\item $\tilde{T}_{R,k}=$max$_j(\tilde{R}_{j,k})$ ($k=1, \ldots, 10^4$). This statistic
picks up the maximum deviation from the continuum spectrum for each simulated PDS. The observed
value $T_R =$max$_j(2MP_j/\hat{S}_j)$ is then compared with the simulated distribution and
the significance is evaluated directly. By construction, it implicitly accounts for the multitrial
search performed all over the frequencies.
\item $A_k$ is the Anderson--Darling (AD) statistic \citep{Anderson52} obtained for the $k$-th set
of $\tilde{R}_{j,k}$ compared with a $\chi^2_{2M}$ distribution.
\item Analogously, KS$_k$ is the Kolmogorov--Smirnov (KS) statistic obtained for the $k$-th set
of $\tilde{R}_{j,k}$ compared with a $\chi^2_{2M}$ distribution.
\end{itemize}
For each of the three statistics, comparing  the values obtained from the observed PDS with the
corresponding distribution of simulated values immediately yields the significance of possible
deviations such as that of a QPO, or the goodness of the fit, as indicated by the AD and KS
statistics. As in G12, in addition to the KS, we chose the AD statistic because it is sensitive
to a few outliers from the expected distribution.

For each GRB the choice between the two competing models was determined by the likelihood ratio
test (LRT) in the Bayesian implementation described by V10. As for the aforementioned statistics,
from the posterior predictive distribution we sampled the pdf of the $T_{LRT}$ statistic
defined as
\begin{eqnarray}
T_{LRT} & = & -\log{\frac{p(\mathbf{P}|\mathbf{\hat{S}_{\rm PL}}, {\rm PL})}{p(\mathbf{P}|\mathbf{\hat{S}_{\rm BPL}}, {\rm BPL})}}\nonumber\\
& = & L(\mathbf{P},\mathbf{\hat{S}_{\rm PL}},{\rm PL}) - L(\mathbf{P},\mathbf{\hat{S}_{\rm BPL}},{\rm BPL})
\qquad ,
\label{eq:T_LRT}
\end{eqnarray}
where $\mathbf{\hat{S}_H}$ denotes the model obtained with the parameters at the mode of the 
posterior distribution of a generic model $H$. The statistics in Eq.~(\ref{eq:T_LRT}) is then sampled
using the simulated PDS $\mathbf{\tilde{P}_k}$ ($k=1,\ldots, 10^4$) and compared with the observed
value. For the LRT test, we performed $10^3$ simulations and accepted the {\sc bpl} model when
the probability of chance improvement was lower than 1\% (see Sect.~\ref{sec:pds_fit}).
As remarked in Sect.~\ref{sec:pds_fit}, while this type of LRT accounts for the whole parameter
space of the simpler model, the same does not hold for the alternative
one, as is the case when we use the Bayes factor. Consequently, the rejection of {\sc pl} does not
necessarily imply that {\sc bpl} is a better option. However, for the limited scope of this investigation,
on independent grounds {\sc bpl} represents a sensible alternative (Sect.~\ref{sec:issue1}).

\begin{figure}
\centering
\includegraphics[width=9.5cm]{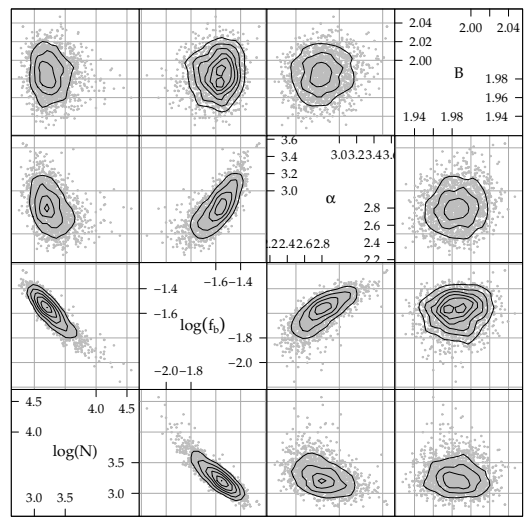}
\caption{Example of marginal posterior distributions for the pairs of parameters
of the {\sc bpl} model obtained from $10^4$ simulated posterior simulations for 050713A (only 2000 points are shown for clarity). Solid lines show the contour levels.}
\label{figA1_neu}
\end{figure}

Finally, after we determined the best--fitting model, we sampled through the MCMC simulation the joint posterior distribution of the model parameters and provided
mean and standard deviation for each of them. As an example, Fig.~\ref{figA1_neu}
shows the marginal posterior distributions of all the pairs of parameters of the {\sc bpl}
model for 050713A.

%
%

\begin{figure}
\centering
\includegraphics[width=8cm]{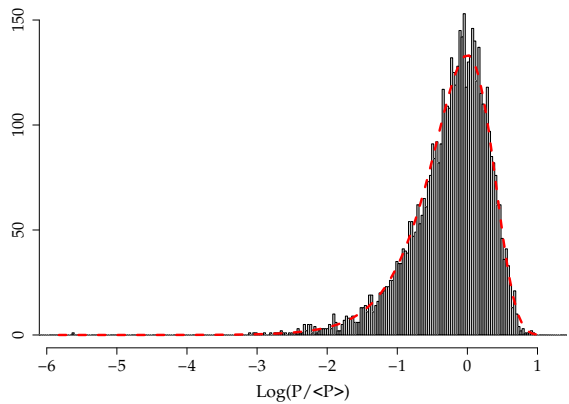}
\caption{Distribution of $\log{(P_j/\hat{S}_j)}$ for 050713A (Fig.~\ref{fig1}).
The dashed line shows the relative renormalised $\chi^2_{2M}$ distribution.}
\label{figA2_neu}
\end{figure}

The goodness of fit is established by the p--values associated to the AD and KS statistics.
In addition, we also compare the distribution of $R_j$ of the observed PDS against the
expected $\chi^2_{2M}$ distribution, as shown in Fig.~\ref{figA2_neu} for 050713A.
%
%

\subsection{Shortcomings of the procedure and implications}
\label{sec:limits_issues}
The procedure relies on the assumption that GRB light curves can be described by stochastic processes resulting
from the convolution of deterministic pulse shape with Poisson point processes. However, this is challenged by
the short-lived and non-stationary nature of GRBs. Specifically, the following assumed properties somehow come into question:
\begin{enumerate}
\item are the {\sc pl} and {\sc bpl} models appropriate to GRBs?
\item  is the unbinned power distributed around the model according to $\chi^2_2$?
\item is the power at different frequency bins distributed in an uncorrelated way?
\end{enumerate}
%

\subsubsection{Are the PDS models appropriate?}
\label{sec:issue1}
Assuming a rough and simplistic pulse shape such as that of a simple exponential with a negligible rise time,
$h(t)=U(t)\,{\rm e}^{-t/\tau}$, where $U(t)$ is the step function ($1$ at $t>0$, $0$ otherwise),
the expected PDS is calculated in a straightforward way,
\begin{equation}
|H(\omega)|^2\ =\ \frac{\tau^2}{1 + (\omega\,\tau)^2}\;,
\label{eq:Pprocess}
\end{equation}
which is included in our {\sc bpl} model for $\alpha=2$ and $\nu_{\rm b}=(2\,\pi\,\tau)^{-1}$.
This description was adopted in the past to estimate the characteristic time of a typical shot in
GRBs \citep{Belli92} and in solar X-ray flares \citep{Frontera79}. Accounting for more
realistic pulse-shape profiles yields complicated, wobbling PDS with a broad range of slopes
at high frequencies \citep{Lazzati02}. This justifies the choice of a general shape like that
of {\sc bpl}, whereas the {\sc pl} is given by the superposition of multiple shots with a range
of different characteristic times as described in Sect.~\ref{sec:res_t90_vs_tau} and
illustrated by D16. An even more general model was succesfully adopted for the PDS of magnetar
outbursts \citep{Huppenkothen13}.

\subsubsection{Is the power $\chi^2_2$-distributed?}
\label{sec:issue2}
We start by considering the results of previous investigations that focused on the average PDS of a sample
of GRBs and on how the power is distributed around it at each given frequency. While the distribution also depends
on the adopted normalisation between different GRBs, the $\chi^2_2$ hypothesis was found to be acceptable
for a peak-count-rate-normalised sample of BATSE GRBs \citep{Beloborodov98} and for a
net-variance-normalised sample of {\em Swift}-BAT GRBs \citep{Guidorzi12}.
The assumption of a GRB as a stochastic process like that of a generalised shot noise convolved with
one or more deterministic pulse shapes provides a theoretical justification for the $\chi^2_2$ distribution of power.
However, in practice the sequence of pulses making up one GRB is limited, and only a few pulses are often observed. As a consequence, the PDS is dominated by the deterministic structure of the few shots (in addition
to the uncorrelated noise that is due to the counting statistics) instead of the stochastic character
brought in by the random point process. In this limit, the power is therefore seen to fluctuate significantly less
than what we would expect from a $\chi^2_2$. This is more evident in the low-frequency bins, where the signal power
dominates the white-noise level \citep{Huppenkothen13}.
This is best illustrated by the limiting case of a single pulse, in which the PDS can be seen as the combination
of a deterministic signal plus uncorrelated noise that is due to counting statistics: in the Leahy normalisation the power
is distributed according to a non-central $\chi^2_2(P_s)$, where $P_s$ is the expected signal power \citep{Groth75,Guidorzi11a}.

This complication is connected with the degeneracy of the problem: we can hardly identify what is due to a deterministic
(typically unknown) signal as opposed to what is due to undersampling (because of the finiteness of the signal itself)
of the stochastic side of the process. This directly brings us to the next question.

\subsubsection{Is the PDS autocorrelated?}
\label{sec:issue3}
In the ideal case of a sufficiently long sequence of pulses, the resulting PDS $\chi^2_2$-oscillates around the average
expected PDS of the deterministic pulses. The PDS of a single pulse can be seen as single realisation of the process,
where the power in each frequency bin is a single realisation of a $\chi^2_2$-distributed variable, whose expected value
is given by the model. However, as noted in the previous section, the PDS of a single pulse fluctuates around the
deterministic model of {\em \textup{that specific}} pulse less than a $\chi^2_2$ at low frequencies. Both properties combined
imply that the PDS of a single pulse presents some degree of autocorrelation: as an example, if the power lies below
the model at a given low frequency where the signal power is much greater than white noise, then a number of
adjacent bins will lie below the model values as well. This undermines the assumption of independence between different
frequency bins upon which the joint likelihood of Eq.~(\ref{eq:jointlik}) is based.

In the attempt of evaluating the effect of this problem on the results, it is worth noting that only the relatively few
low-frequency bins are affected. On the other hand, this is precisely the frequency range where the comparison between
the two competing models mostly matters. In addition, it is reassuring that the preference for {\sc bpl} over {\sc pl},
as established by a likelihood that overestimates the null-hypothesis variance, works conservatively so that
{\sc bpl} cannot be mistakenly preferred and is better a fortiori.
%
\begin{figure}
\centering
\includegraphics[width=8cm]{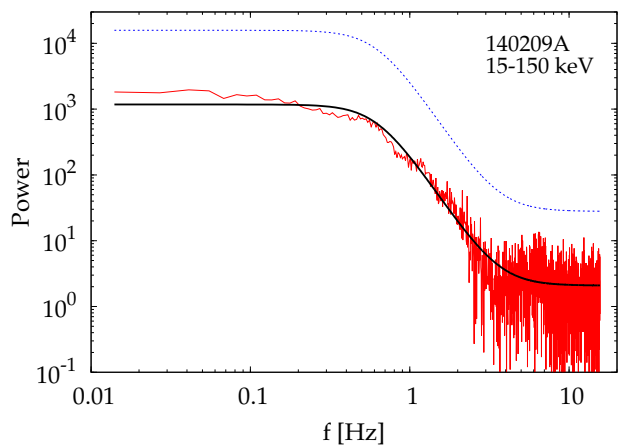}
\caption{PDS and best-fit model of 140209A, the GRB with the lowest p-value of the preferred model (a {\sc bpl}).
The poor quality is mainly due to the low-frequency correlated power, which oscillates less than a $\chi^2_2$ around
the model. Nonetheless, the overall shape of the PDS is well described.}
\label{figA3_neu}
\end{figure}
%
As an illustrative case, we show the PDS of 140209A (Fig.~\ref{figA3_neu}), which has the lowest p-value of the best-fit models of all our samples (Sect.~\ref{sec:res}).
The quality of the fit is very poor, as assessed by the compatibility with the assumption that
the PDS oscillates around the model as a $\chi^2_2$. However, although the uncertainties on the best-fit parameters are
to be taken with caution, the overall shape of the PDS is clearly very well described by the preferred model ({\sc bpl}
in this case).

To summarise, we conclude that despite the problems at low-frequencies, the assumptions upon which our procedure relies are
acceptable and the breakdown of the statistical independence of the power at low frequencies does not alter its
conservative character as for the preference for the more complicate model of {\sc bpl} over the simple {\sc pl}.


\end{document}